\documentclass[fleqn,usenatbib]{mnras}

\usepackage{newtxtext,newtxmath}

\usepackage[T1]{fontenc}

\DeclareRobustCommand{\VAN}[3]{#2}
\let\VANthebibliography\thebibliography
\def\thebibliography{\DeclareRobustCommand{\VAN}[3]{##3}\VANthebibliography}

\usepackage{graphicx}	
\usepackage{amsmath}	
\usepackage{array,multirow}
\usepackage{tabularray}
\usepackage{subcaption}

\newcommand{\eazy}{{\tt{EAZY-py}}}

\newcommand{\bagpipes}{{\textsc{bagpipes}}}

\newcommand{\prospector}{{\textsc{prospector}}}
\newcommand{\sextractor}{{\tt{SExtractor}}}

\newcommand{\galfind}{{\tt{GALFIND}}}

\newcommand{\cloudy}{{\textsc{cloudy}}}

\def\casgm20{CAS-G-M$_{20}\,$}
\def\m20{M$_{20}\,$}

\newcommand{\synthesizer}{\textsc{synthesizer}}
\newcommand{\synference}{\textsc{synference}}
\newcommand{\ltuili}{\textsc{LtU-ILI}}
\newcommand{\modelone}{\texttt{Model\,1}}
\newcommand{\modeltwo}{\texttt{Model\,2}}

\title[Synference]{Flexible Simulation Based Inference for Galaxy Photometric Fitting with Synthesizer}

\author[Harvey et al.]{
Thomas Harvey$^{1}$\thanks{E-mail: thomas.harvey-3@manchester.ac.uk},
Christopher C. Lovell$^{2,3}$,
Sophie Newman$^{4}$,
Christopher J. Conselice$^{1}$,
Duncan Austin$^{1}$,\newauthor
William J. Roper$^{5}$,
Aswin P. Vijayan$^{5}$,
Stephen M. Wilkins$^{5}$,
Patricia Iglesias-Navarro$^{6, 7}$,
Vadim Rusakov$^{1}$,  \newauthor
Qiong Li$^{1}$,
Nathan Adams$^{1}$,
Kai Magdwick$^{1}$,
Caio M. Goolsby$^{1}$,
Marc Huertas-Company$^{6, 7, 8, 9}$,
and Matthew Ho$^{10}$ \\
$^{1}$Jodrell Bank Centre for Astrophysics, University of Manchester, Oxford Road, Manchester M13 9PL, UK\\
$^{2}$Kavli Institute for Cosmology, Madingley Road, Cambridge CB3 0HA, UK \\
$^{3}$Institute of Astronomy, Madingley Road, Cambridge CB3 0HA, UK\\
$^{4}$Institute of Cosmology and Gravitation, University of Portsmouth
, Burnaby Road, Portsmouth PO1 3FX, UK\\
$^{5}$Astronomy Centre, University of Sussex, Falmer, Brighton BN1 9QH, UK\\
$^{6}$Instituto de Astrofísica de Canarias, C/ Vía Láctea s/n, 38205 La Laguna, Tenerife, Spain\\
$^{7}$Departamento de Astrofísica, Universidad de La Laguna, 38200 La Laguna, Tenerife, Spain\\
$^{8}$Observatoire de Paris, LERMA, PSL University, 61 avenue de l’Observatoire, F-75014 Paris, France\\
$^{9}$Université Paris-Cité, 5 Rue Thomas Mann, 75014 Paris, France\\
$^{10}$Department of Astronomy, Columbia University, New York, NY 10027, USA\\}
\date{Accepted XXX. Received YYY; in original form ZZZ}
\pubyear{\the\year{}}

\begin{document}
\label{firstpage}
\pagerange{\pageref{firstpage}--\pageref{lastpage}}
\maketitle

\begin{abstract}
We introduce \synference{}, a new, flexible Python framework for galaxy SED fitting using simulation-based inference (SBI). \synference{} leverages the \synthesizer{} package for flexible forward-modelling of galaxy SEDs and integrates the \ltuili{} package to ensure best practices in model training and validation. In this work we demonstrate \synference{} by training a neural posterior estimator on $10^6$ simulated galaxies, based on a flexible 8-parameter physical model, to infer galaxy properties from 14-band HST and JWST photometry. We validate this model, demonstrating excellent parameter recovery (e.g. R$^2>$0.99 for M$_\star$) and accurate posterior calibration against nested sampling results. We apply our trained model to 3,088 spectroscopically-confirmed galaxies in the JADES GOODS-South field. The amortized inference is exceptionally fast, having nearly fixed cost per posterior evaluation and processing the entire sample in $\sim$3 minutes on a single CPU (18 galaxies/CPU/sec), a $\sim$1700$\times$ speedup over traditional nested sampling or MCMC techniques. We demonstrate \synference{}'s ability to simultaneously infer photometric redshifts and physical parameters, and highlight its utility for rapid Bayesian model comparison by demonstrating systematic stellar mass differences between two commonly used stellar population synthesis models. \synference{} is a powerful, scalable tool poised to maximise the scientific return of next-generation galaxy surveys.
\end{abstract}

\begin{keywords}
methods: data analysis -- techniques: photometric -- galaxies: photometry -- galaxies: stellar content
\end{keywords}



\section{Introduction}

Current and future space and ground-based surveys by missions including the James Webb Space Telescope \citep{JWST_overview}, Euclid \citep{Euclid_overview}, the Nancy Grace Roman Space Telescope \citep{WFIRST_overview}, the Vera C. Rubin Observatory \citep{LSST_overview} and SPHERE-X \citep{SPHERE-X_overview}, among others, will cumulatively observe $>20$ billion galaxies in the next decade. The unprecedented precision and volume of this data necessitate the development of increasingly sophisticated and performant models to interpret the diverse range of observed galaxy properties. Constraining these models is therefore a critical step towards a more complete understanding of galaxy formation and evolution. However, conventional Bayesian techniques for posterior parameter inference, such as Markov Chain Monte Carlo or nested sampling, are computationally prohibitive for such large-scale analyses. Inference for a single galaxy can require anywhere from minutes \citep[e.g. \textsc{bagpipes};][]{carnallInferringStarformationHistories2017} to several days \citep[e.g. \textsc{prospector};][]{johnsonStellarPopulationInference2020, tacchellaStellarPopulationsGalaxies2022} of CPU time, depending on model complexity and forward model efficiency, rendering these methods orders of magnitude too slow for these forthcoming datasets. As a result, there have been numerous approaches to improving the speed of SED fitting using a variety of machine learning methods. One approach is to emulate the simulator using a neural network emulator \citep{alsing2020speculator, mathews2023simple}, whilst other approaches have utilized approaches including manifold learning \citep{2019ApJ...881L..14H,2022A&A...665A..34D,2025ApJ...989...65A}, de-noising autoencoders \citep{2017A&A...603A..60F}, Convolutional Neural Networks \citep{lovell2019learning,dobbels2019morphology,surana2020predicting}, and Natural Gradient Boosting \citep{gilda2021mirkwood}. These approaches typically produce only a point parameter estimate, rather than a full posterior, and can be quite sensitive to the training dataset used.

A promising solution to this scalability challenge is \textit{simulation-based inference} (SBI), also known as \textit{likelihood-free inference} and \textit{implicit likelihood inference} \citep{cranmer2020frontier}. SBI methods circumvent the need for an explicit likelihood function, Instead learning a statistical mapping between observations and data based on Bayesian statistics. The SBI approach utilizes a neural network as a Neural Density Estimator (NDE), which is trained to predict the posterior (Neural Posterior Estimation), likelihood (Neural Likelihood Estimation), or the likelihood-to-evidence ratio (Neural Ratio Estimation). A key advantage of SBI, and NPE in particular, is its capacity for amortized posterior estimation. Once a neural network is trained on a comprehensive set of simulations, it can generate posterior distributions for new observations almost instantaneously, without requiring any further optimization. This amortized nature enables efficient scaling to the vast datasets produced by upcoming surveys. Furthermore, the resulting posterior distributions capture the full uncertainty landscape, including parameter degeneracies, which is a significant advantage over optimisation methods that yield only point estimates, such as $\chi^2$ minimisation.

In recent years, SBI has been successfully applied across a range of astrophysical and cosmological contexts. For individual galaxies, applications include spectral energy distribution (SED) fitting with photometry/spectroscopy \citep{hahnAcceleratedBayesianSED2022, 2022MLS&T...3dLT04K, wangSBIFlexibleUltrafast2023, fischbachergalsbi2024}, star formation history (SFH) reconstruction from photometry and spectroscopy \citep{aufortReconstructingGalaxyStar2024, iglesias-navarroDerivingStarFormation2024}, decomposition of active galactic nuclei (AGN) from their host galaxies \citep{bergerBiasesStellarMasses2025, yuanDisentanglingStarFormation2025}, and inference of galaxy dynamics \citep{widmarkUsingSimulationBased2025, 2025MNRAS.542.1776S, 2025arXiv251004735B}. The technique has also been extended to spatially-resolved SED fitting \citep{iglesias-navarroSimulationbasedInferenceGalaxy2025}. On larger scales, SBI has been used to infer dark matter halo properties \citep{hahnHaloFlowNeuralInference2024}, the history of cosmic reionization \citep{chenLearningReionizationHistory2023, choustikovInferringIonizingPhoton2025}, and high-dimensional parameters of population-level models \citep{lovell2024learning, 2024JCAP...05..049M,2024AJ....167...16L,thorpPopcosmosScaleableInference2024,thorpPopcosmosInsightsGenerative2025, gunes2025compass, 2025arXiv250920430D}. SBI is also used extensively within cosmology for parameter inference \citep[e.g.][]{alsing2019fast, modiSensitivityAnalysisSimulationBased2023, mediato2025cosmological, zubeldia2025extracting, 2025MNRAS.536.1303J}.

In this work, we present \synference{}, a new, flexible SBI framework designed for robust galaxy SED fitting. \synference{} leverages the \synthesizer{} package \citep{lovell2025synthesizer, roper2025synthesizer} for the flexible creation of training data and integrates the \ltuili{} package \citep{ho2024ltu} to ensure best practices in model configuration, training, and validation. To demonstrate its capabilities, we apply \synference{} to infer the physical properties of galaxies from the JWST Advanced Deep Extragalactic Survey  \citep[JADES, ][]{eisensteinOverviewJWSTAdvanced2023, riekeJADESInitialData2023}, using a combination of JWST and Hubble Space Telescope (HST) photometry. 

The paper is structured as follows. In \autoref{sec:sbi_bckg}, we provide a technical overview of SBI methods. In \autoref{sec:synference} we provide an overview of the \synference{} framework. \autoref{sec:training} details the generation of our training data and the model training process, followed by a thorough evaluation of the trained model in \autoref{sec:validation}. We apply our framework to the JADES dataset in \autoref{sec:applications}. In \autoref{sec:discussion}, we discuss the limitations of our model, compare our results with those in the literature, and outline future directions. Finally, we summarize our findings in \autoref{sec:conclusions}.

Throughout this paper, we assume a standard $\Lambda\mathrm{CDM}$ cosmology with $H_0=70$\,km\,s$^{-1}$\,Mpc$^{-1}$, $\Omega_{\rm M}=0.3$ and $\Omega_{\Lambda} = 0.7$.

\section{Simulation-based Inference}
\label{sec:sbi_bckg}
This section briefly reviews key concepts of simulation-based inference. SBI is a class of machine learning techniques for performing parameter inference in scenarios where the underlying physical process can be simulated but the likelihood function is analytically intractable or computationally prohibitive to evaluate. In many scientific domains, it is straightforward to construct a stochastic simulator that executes the "forward problem": predicting an observation $x$ given a set of model parameters $\theta$. However, scientific inquiry typically requires solving the "inverse problem": inferring the posterior probability distribution of the parameters given an observation, $p(\theta|x_i)$,

According to Bayes' theorem,  the posterior is given by:
\begin{equation}
    p(\theta|x_i) = \frac{p(x_i|\theta)p(\theta)}{p(x_i)}
\end{equation}

\noindent where $p(\theta)$ is the \textit{prior} distribution of the parameters, $p(x_i|\theta)$ is the \textit{likelihood} and $p(x_i)$ is the \textit{marginal likelihood} or \textit{evidence}.
SBI methods bypass the need for an explicit likelihood function by instead learning the statistical relationship between parameters and observations from a training set of simulated observation-parameter pairs, $\mathcal{D} = \{(x_i, \theta_i)\}^N_{i=1}$.  

To achieve this, SBI leverages a technique called neural density estimation \citep{papamakariosMaskedAutoregressiveFlow2018}, where a neural network is trained to approximate a conditional probability distribution. Depending on which component of Bayes' theorem is approximated, SBI methods can be broadly categorized.

\subsection{Neural Posterior Estimation}

In Neural Posterior Estimation (NPE), the network directly learns the posterior distribution  $p(\theta|x)$ \citep{greenbergAutomaticPosteriorTransformation2019}. This is achieved by training a conditional probability distribution,  $q_\phi(\theta|x)$, (where $\phi$ represents the network weights) to approximate the true posterior. The training process typically minimizes the Kullback-Leibler (KL) divergence \citep{kullback1951information} between the learned and true posteriors, averaged over all simulated data. 

\begin{equation}
    D_{\rm KL}(p||q) = \int p(\theta|x) \log\frac{p(\theta|x)}{q_\phi(\theta|x)}d\theta
\end{equation}

We can define a loss function $L$ by taking the expectation $\mathbb{E}_{p(x)}$ over the training dataset $\mathcal{D}$, for which an optimizer \citep[e.g. Adam][]{kingma2014adam} is used to minimize the gradient of $L$ with respect to the network parameters.

\begin{equation}
    \mathcal{L} = \mathbb{E}_{p(x)}D_{KL} = \mathbb{E}_{p(x, \theta)}[- \log q_\phi(\theta|x)]
\end{equation}
In practice, this loss function encourages the network to maximize the probability of the true parameters $\theta_i$ that correspond to each simulated observation $x_i$ in the training set.


A large number of different model architectures have been used to represent the conditional probability distribution $q(\theta|x)$. There are two major categories which are commonly used; \textbf{normalising flow} \citep[NF, ][]{kobyzevNormalizingFlowsIntroduction2021, papamakarios2021normalizing} models and \textbf{mixture density networks} \citep[MDN, ][]{bishop1994mixture}. MDNs are a neural architecture where the conditional probability distribution is represented by a mixture model with learned weights and biases. Normalising flow architectures, such as \textbf{Masked Autoregressive Flows} \citep{papamakariosMaskedAutoregressiveFlow2018} and \textbf{Neural Spine Flows} \citep{papamakarios2021normalizing}, use invertible transformations of a base Gaussian distribution to represent complex conditional probability distributions using a neural network. Alternative approaches, such as using transformer-based diffusion models, have also shown promising performance \citep{thorpPopcosmosScaleableInference2024,thorpPopcosmosInsightsGenerative2025,gloecklerAllinoneSimulationbasedInference2024}.

Other SBI families include Neural Likelihood Estimation (NLE) \citep{papamakariosSequentialNeuralLikelihood2019}, which learns the likelihood $p(x|\theta)$ and Neural Ratio Estimation (NRE), which learns a likelihood ratio \citep{hermansLikelihoodfreeMCMCAmortized2020, durkanContrastiveLearningLikelihoodfree2020}. While powerful, both NLE and NRE require an additional sampling step (e.g., MCMC) to generate posteriors for a given observation, making them less computationally performant than NPE for the large-scale applications motivating this work. Therefore, we primarily focus on the NPE method in this study, although \synference{} does support NLE and NRE, as well as the sequential variants (SNPE, SNLE, SNRE), where multiple rounds of training are performed.

\subsection{Strengths of SBI}

SBI, and particularly NPE, offers several key advantages over traditional inference techniques:

\begin{enumerate}
    \item \textbf{Amortized Inference:} Once an NPE model is trained, generating posterior samples for a new observation is computationally trivial. This "amortization" of the inference cost across all possible observations means that analysing millions of galaxies becomes feasible, as the expensive training phase is performed only once.
    \item \textbf{Full Posterior Characterization:} Unlike optimisation methods that yield only a single best-fit point estimate (e.g., a maximum likelihood value), SBI provides the full posterior distribution. This captures all parameter uncertainties and degeneracies, providing a more complete and robust picture of the inference result.
    \item \textbf{Likelihood-Free:} SBI is applicable to any scientific model from which one can forward-simulate data. This flexibility allows for the use of highly complex and realistic simulators for which an analytical likelihood is unknown. In the context of SED fitting this allows utilization of e.g. forward-modelled hydrodynamical simulation outputs as the training dataset directly \citep[e.g.][]{choustikovInferringIonizingPhoton2025}. 
    \item \textbf{Model Comparison:} SBI techniques can be leveraged for Bayesian model comparison either by learning the posterior and likelihood \citep{spuriomanciniBayesianModelComparison2023} or directly training a neural network classifier to predict the Bayes factor between two competing models \citep{jeffreyEvidenceNetworksSimple2024}. Alongside the speed of inference these techniques make SBI a competitive tool for Bayesian model selection.
\end{enumerate}

\subsection{Practical Considerations and Challenges}

While powerful, the application of SBI requires careful consideration of its implementation, potential challenges, and validation. In this section we briefly outline and discuss these challenges in the context of SED fitting. 

\subsubsection{SBI Implementation}

A number of open-source Python packages have been developed to streamline the implementation of SBI. Notable examples include \textsc{sbi} \citep{tejero2020sbi}, \textsc{lampe} \citep{boelts2024sbi}, and \textsc{pydelfi}  \citep{alsing2019fast}, as well as newer approaches (e.g. \textsc{simformer}) implementing novel architectures such as transformers \citep{gloecklerAllinoneSimulationbasedInference2024}. These packages are built on top of the machine learning and GPU acceleration frameworks \textsc{PyTorch, tensorflow} and \textsc{jax} \citep{paszke2019pytorch, developers2022tensorflow, bradbury2021jax}. The \textsc{LtU-ILI} package \citep{ho2024ltu} further unifies several of these tools into a cohesive framework and implements consistent best practices for model evaluation. In this paper we use LtU-ILI with the \textit{sbi} backend for our fiducial model.

\subsubsection{Ongoing Challenges}

Several challenges must be addressed when applying SBI to astrophysical problems like SED fitting.

\begin{itemize}
    \item \textbf{Expensive Forward Modelling:} SBI relies on a library of observation parameter pairs generated through a forward model. Depending on the complexity of the forward model or simulation this can be a significant burden. However emulators which predict the observations given the parameters have also been successfully used to build training datasets for SBI, e.g. \textsc{speculator} \citep{alsing2020speculator} or Parrot \citep{mathews2023simple}. Active learning \citep{griesemerActiveSequentialPosterior2024} can also significantly reduce the required library size for specific applications. 
    \item\textbf{Designing a training dataset:} The explicit and implicit prior choices in the training dataset will be learned by the model, and these need to be chosen carefully to ensure the model is both applicable and unbiased, as well as precise enough for the intended application. Broad parameter sampling is important for accurate results, and low-dispersion sampling methods such as Latin Hypercube or Sobol Sequence can ensure this whilst reducing the total number of simulations required. The size of the training set required typically scales with the dimensionality of the inference \citep{bairagi2025many}, and typically ranges from $10^{4}$ to $10^6$ for parameter inference with 5-10 dimensions. 
    \item \textbf{Model misspecification:} This occurs when the simulator used for training is an imperfect representation of reality, or when an observation lies outside the parameter space covered by the training data \citep{cannonInvestigatingImpactModel2022}. Applying a trained model to such "out-of-distribution" data can lead to biased and unreliable posteriors. Methods to detect misspecification include checking for consistency across an ensemble of models \citep{alveySimulationbasedInferenceDeep2025}, using outlier detection algorithms as a prior predictive check \citep{aufortReconstructingGalaxyStar2024}, or explicitly modelling potential mismatches, as demonstrated by \textsc{sbi++} \citep{wangSBIFlexibleUltrafast2023}. Techniques such as Domain Adaption have also been used to enable neural networks to trained on simulated data perform more accurately when applied to real data \citep[e.g. for SDSS galaxies trained on Illustris simulations in ][]{2021MNRAS.506..677C}.
    \item\textbf{Noise and Uncertainty Modelling:} If we wish to apply our trained model to reality, we must model the imperfect measurement process which introduces both systematic and random noise into our observations. Modelling this noise accurately is important for the trained SBI model to perform accurately when applied to real data. 
    In astrophysical observations we often deal with heteroscedastic, non-Gaussian and correlated noise which can be difficult to model, and often differs between different observations and surveys. Various techniques to model noise have been suggested, including de-noising score matching \citep{vincent2011connection}, noise-insensitive loss functions (Huber loss) , or direct modelling of noise distributions using a trained score-based diffusion model \cite[e.g.][]{thorpPopcosmosInsightsGenerative2025}.
    \item \textbf{Missing Data:} Heterogeneous datasets are common in astronomy, such as missing or incomplete photometric survey coverage in certain filters. Standard 'SBI' approaches, like many ML techniques, require a fixed input data vector of known size. Strategies to handle this include imputing missing values using nearest-neighbour searches in the training set \citep{wangMonteCarloTechniques2022, wangSBIFlexibleUltrafast2023}, or conditioning the model on the data availability pattern itself by using a binary mask as an additional input \citep{wangMissingDataAmortized2024}. Architectures that use tokenization and attention mechanisms, such as transformers, are also naturally suited to handling missing data \citep{gloecklerAllinoneSimulationbasedInference2024}.
    \item \textbf{Sensitivity to Hyperparameters:} The computational and posterior predictive performance of an SBI model is sensitive to choices of neural architecture, network size, learning rate, and other training hyperparameters. Finding the optimal configuration often involves navigating a bias-variance trade-off. This process can be automated using optimization frameworks like \textsc{Optuna} \citep{akiba2019optuna}. Additionally, creating an ensemble of models trained with different hyperparameters can improve robustness and provide more reliable posterior estimates \citep{hermans2021trust}, at the cost of more computation.
\end{itemize}

\subsubsection{Benchmarking SBI Models}
\label{sec:benchmark}
Once the SBI model has been trained, validating model performance is critical to understanding whether the model is accurate and unbiased before it is applied to real observations. This benchmarking should always be performed on a held-out test set to ensure the model generalizes well and is not overfit to the training data \citep{lueckmannBenchmarkingSimulationBasedInference2021}. Commonly used benchmarks for precision include comparison of true vs predicted marginal parameters, producing corner plots showing multivariate distributions of parameter pairs, or the cumulative likelihood of the validation dataset \citep{ho2024ltu}.

A model can be precise but biased, or it may underestimate the true uncertainty in the posterior distribution. If available, tests against reference posteriors (obtained by nested sampling or MCMC) such as Classifier 2 Sample Tests \citep[C2ST, ][]{lopez2016revisiting} and Probability-Integral Transform (PIT)/Simulation-based calibration (SBC) \citep{cook2006validation, zhao2021diagnostics, talts2018validating} can be used to assess the performance of a model. However reference posteriors are typically computationally expensive, so alternative methods like coverage tests, \textit{Tests of Accuracy with Random Points} \citep[TARP, ][]{lemos2023sampling}, truncated proposals \cite{deistlerTruncatedProposalsScalable} can be used to evaluate multivariate and marginal posterior coverage.

\section{\synference{}: A flexible SBI Framework for SED Fitting}
\label{sec:synference}

To address the aforementioned issues prevalent in SBI in the context of SED fitting, we introduce \synference{}, a Python package that provides a flexible framework for building, training, and applying SBI models to infer galaxy properties from photometric and spectroscopic observations. With \synference{} we aim to make utilizing SBI approaches for SED fitting more accessible to users without ML experience. 

\synference{} is designed to provide flexibility to the user in every stage, which we have conceptually split into distinct stages:

\begin{enumerate}
    \item \textbf{Simulation}: A forward model is used to generate a comprehensive library of simulated observations ($x$) and their corresponding physical parameters ($\theta$). By default, \synference{} uses \synthesizer{} for this stage, but it is also compatible with models from other SED modelling codes (e.g., \bagpipes{}, \prospector{}) or forward modelled emission from cosmological models, e.g.  hydrodynamic simulations such as EAGLE \citep{schaye2015eagle, crain2015eagle} and IllustrisTNG \citep{pillepich2018simulating}). More generally, any simulator which predicts astrophysical observables given some parameters, for example output emission from radiative transfer modelling codes \citep[e.g., SKIRT ][]{camps2015skirt}, Stellar Population Synthesis (SPS) codes  \citep[e.g., FSPS, BPASS, ][]{conroy2010fsps,stanway2018re}), photoionization modelling codes \citep[\cloudy, ][]{2017RMxAA..53..385F} can all in theory used to train a model using  \synference{}.
    \item \textbf{Feature Engineering}: The raw simulation library is processed to create a training-ready dataset. This involves selecting specific observables (e.g., photometric filters, wavelength range/resolution), applying realistic noise models, and performing data transformations to ensure numerical stability during training.
    \item \textbf{Training and Validation}: An SBI model is trained on the engineered feature set. Through its \ltuili{} backend, \synference{} supports a wide variety of neural density estimators for NPE, NLE, and NRE, with hyperparameter optimisation provided by \textsc{Optuna}. Model performance and validation are run by default for every model (e.g. coverage plots, rank histogram, parameter metrics, TARP, training loss, multivariate corner plots for posteriors, and log probability distributions). This makes it easy for the end-user to assess the performance and applicability of their trained model. 
    \item \textbf{Inference}: The trained model is applied to real observational data. This stage includes automated data transformations, prior predictive checks to flag model misspecification, and built-in tools for posterior visualization.
\end{enumerate}

\begin{figure*}
    \centering
    \includegraphics[width=0.9\textwidth]{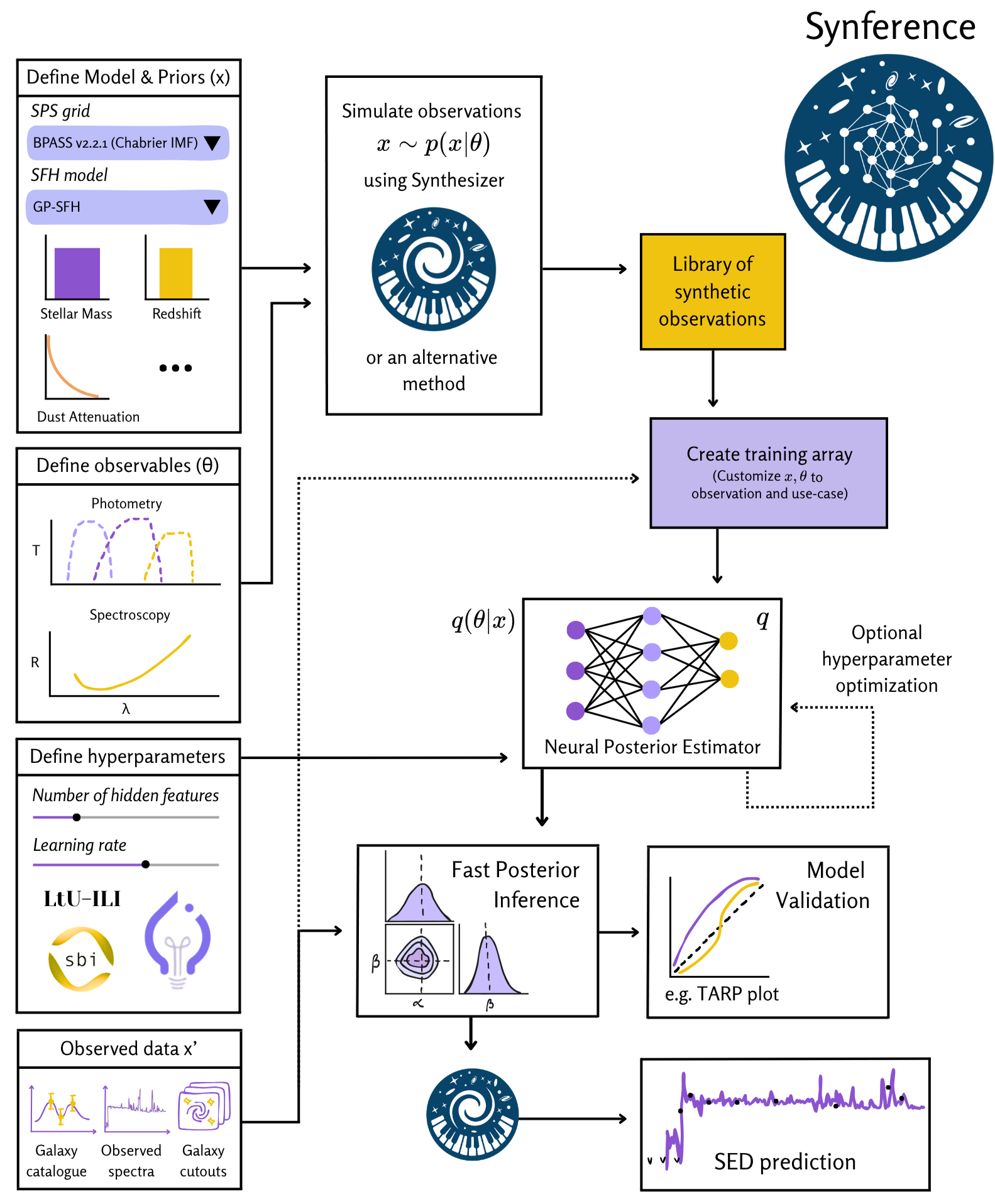}
    \caption{Schematic illustration of the \synference{} workflow. User-defined inputs are shown on the left, including the physical model configuration within \synthesizer{}, the choice of instrumentation (photometric filters, wavelength range, spectral resolution), as well as the SBI model hyperparameters for the neural density estimator, which utilizes the \textsc{ltu-ili} package \citep{ho2024ltu} with either the \textsc{sbi} \citep{tejero2020sbi} or \textsc{lampe} \citep{boelts2024sbi} backends. The \synference{} framework processes these to generate a library of mock observations, either using the \synthesizer{} package or directly from a user-provided model. From this library we generate a training array for the SBI model, then train a neural density estimator to emulate the posterior or likelihood. The hyperparameters can be optionally optimized using \textsc{optuna} \citep{akiba2019optuna}. The amortized trained model can then be utilized for direct inference from observed data to generate reliable Bayesian posterior estimates. Optionally these posteriors can be used to reconstruct the predicted SED of the galaxy.}
    \label{fig:synference}
\end{figure*}

A key design principle of \synference{} is the decoupling of the simulation and training stages. This modularity allows a single, comprehensive library of simulated SEDs to be reused for training multiple SBI models, each tailored to a specific observational dataset or scientific question. The framework is built for scalability, with MPI support for parallelized data generation on high-performance computing (HPC) systems. \synference{} is publicly available on GitHub\footnote{\url{https://github.com/synthesizer-project/synference}}. A schematic of the overall workflow is shown in \autoref{fig:synference}, highlighting the process of constructing, building and training a model using \synference{}.

\subsection{The Forward Model: \synthesizer{}}
\label{sec:simulator}

The foundation of any SBI application is a high-fidelity forward model. \synference{} leverages \synthesizer{} \citep{lovell2025synthesizer, roper2025synthesizer}, a Python package for generating synthetic galaxy observations. \synthesizer{} can produce complex, multi-component SEDs incorporating stellar continuum, nebular emission (using the photoionization code \textsc{cloudy}, \citealt{2017RMxAA..53..385F}), dust attenuation and emission, AGN emission, and intergalactic medium (IGM) absorption. It supports a wide range of SPS models, initial mass functions (IMFs), star formation histories, and dust laws, as summarized in \autoref{tab:model_params}. This flexibility allows users to construct bespoke physical models tailored to their specific scientific goals. One of the unique approaches of \synthesizer{} is the unified approach to forward modelling from both parametric models and outputs of hydrodynamical simulations, allowing like-for-like comparisons between inverse and forward modelling approaches.

Throughout the rest of this paper we adopt a number of fiducial modelling choices for the purposes of demonstrating the \synference{} framework, which we call \modelone{}. We model the stellar emission using the BPASS v2.2.1 SPS grids \citep{stanway2018re}, assuming a \cite{chabrierGalacticStellarSubstellar2003} IMF with mass limits of 0.1 and 300$M_\odot$. Nebular emission is included assuming a fixed ionization parameter of $\log_{10}U=-2$. Dust attenuation is modelled with a single-component \cite{calzetti2000dust} and dust emission is modelled as a greybody with a fixed temperature of $40$K and emissivity $\beta=1.5$, with IR flux density calculated assuming overall energy balance with reprocessed ionizing photons. IGM attenuation is applied using the model of \cite{Inoue2014}.

Whilst we make these explicit choices in order to demonstrate \synference{} through \modelone{}, these decisions can be fully configured by the end user, and \synthesizer{} offers a very wide variety of options for different use-cases. 

On the observation side \synthesizer{} supports all instruments provided by the Spanish Virtual Observatory \citep[SVO;][]{rodrigo_svo_2012,rodrigo_svo_2020}, as well as custom filter profiles, so photometry can be generated for any combination of filters. \synference{} also supports inference directly from spectral observations, given that an embedding network is provided to reduce the dimensionality of the observations. Whilst detailed applications to spectroscopy are beyond the scope of this work, we plan  to explore this functionality of \synference{} in future work. 

\begin{table}

\caption{An overview of the grids and parametric modelling components available in \synthesizer{}.}
\label{tab:model_params}

\resizebox{1 \columnwidth}{!}{
\begin{tabular}{l}
\multicolumn{1}{c}{\textbf{SPS Grids}}                                                                                                                               \\ \hline
BC03 \citep{bruzualStellarPopulationSynthesis2003}, BPASS \citep{stanway2018re}, \\ FSPS \citep{conroy2010fsps}, Maraston \citep{maraston2005evolutionary,2025MNRAS.tmp.1769N}, \\ Yggdrasil \citep{zackrissonSPECTRALEVOLUTIONFIRST2011}                                                    \\
\\
\multicolumn{1}{c}{\textbf{Star Formation Histories}}                                                                                                                \\ \hline
\textbf{Parametric: }Constant, Exponential, Declining Exponential, \\

Delayed Exponential, Lognormal, Double Powerlaw. \\

\textbf{Non-parametric} "Continuity", "Continuity Flex" \citep{lejaHowMeasureGalaxy2019}, \\ "Continuity PSB" \citep{suess2022recovering}, \\ "Dirichlet" \citep{leja2017deriving, lejaHowMeasureGalaxy2019}, Dense Basis \citep{iyerNonparametricStarFormation2019},  \\ 

\\

\multicolumn{1}{c}{\textbf{Dust Attenuation and Emission}}                                                        \\ \hline
\begin{tabular}[c]{@{}l@{}}\textbf{Attenuation Laws}: Powerlaw, \cite{calzetti2000dust}, \\\cite{charlot2000simple}, \cite{li2008dust} [MW, SMC, LMC]. \\ \textbf{Emission}: Blackbody, Greybody, \cite{casey2012far}, \cite{draine2007infrared}\end{tabular} \\
\textbf{IGM Attenuation}: \cite{Madau1995, Inoue2014, asada2025improving} \\
\\

\multicolumn{1}{c}{\textbf{AGN Modelling}}      \\
\hline

\textsc{UnifiedAGN}: QSOSED \citep{KD18} model [Torus, NLR, BLR]

\end{tabular}}
\end{table}

\subsubsection{Star Formation History Parametrizations}

A significant number of star-formation history parametrizations have been suggested in the literature to represent the complex and diverse growth of stellar populations. SFH timescales can vary significantly between galaxies, from stochastic bursts on Myr timescales in low-mass galaxies to smoothly varying across Gyr timescales in massive quiescent galaxies \citep{iyer2020diversity}. Representing these diverse shapes and timescales of variability with a simple model is an ongoing challenge.

Proposed models broadly fall into two groups; `parametric' models, typically represented by smoothly time-varying functions such as a `delayed' or `declining' exponential, or a more complex lognormal or double power-law model \citep{carnallHowMeasureGalaxy2019}, and `non-parametric' models, where a SFH is represented by a time-binned or stochastic process, such as the `continuity' model of \cite{lejaHowMeasureGalaxy2019,tacchellaStellarPopulationsGalaxies2022}, the `Dense Basis' model of \cite{iyerNonparametricStarFormation2019}, or the stochastic SFH power-spectrum prior of \cite{2024MNRAS.532.4002W} \citep[see also][]{2024ApJ...961...53I}. The `non-parametric' models typically have more degrees of freedom than parametric models but may better represent the diversity of observed SFHs, although care must be taken to select appropriate hyperparameters and priors for these models.

Through \synthesizer{}, \synference{} supports many common parametric and `non-parametric' SFHs, as listed in \autoref{tab:model_params}, and users can also simply add additional models as desired. 
For \modelone{} we implement a flexible Gaussian-process-based SFH from \cite{iyerNonparametricStarFormation2019}, where the SFH is controlled by a decoupled recent SFR and $N=3$ parameters controlling the lookback time at which 25 / 50 / 75\% of the total stellar mass formed (t$_{25}$, t$_{50}$, t$_{75}$). The prior on these parameters takes the form of a Dirichlet distribution, for which we set $\alpha=1$, which controls the correlation of the parameters and the condition $t_{25} < t_{50} < t_{75}$. Whilst increasing the number of parameters would allow for more complex star formation histories, it also becomes more difficult for the neural network to recover the true star formation history, particularly when attempting to infer it from photometric observations.

\subsection{Simulating mock observations}

We list our prior parameters and their allowed ranges for \modelone{} in  \autoref{tab:priors}. Using the \modelone{} simulator, we generate a large training dataset by sampling from the prior distributions defined in \autoref{tab:priors}. We adopt broad, non-informative uniform or log-uniform priors for our free parameters to ensure the training set covers a wide hypervolume of galaxy properties. We impose a sSFR prior to cover a range of sSFR values, but we choose to infer $\log_{10}$ SFR directly rather than the specific star formation rate. The free parameters in \modelone{} are: stellar mass $\rm M_\star$, V-band dust attenuation ($\rm A_V$), stellar metallicity ($\rm Z_\star$), star formation rate (SFR), and the three SFH parameters ($t_{25},t_{50},t_{75}$), for a total of seven free parameters. For each simulated observation we also calculate several derived parameters which we can train the model to infer directly, including mass-weighted age, star-formation rates averaged over several timescales, UV $\beta$ slope and the surviving stellar mass calculated from the SPS return fractions given the SFH and metallicity. 

\begin{table*}
    \caption{Free and fixed parameter values and priors for \protect\modelone{}. The upper table shows the varying parameters and their associated prior distributions, along with a brief description. The lower table shows the fixed values for other parameters assumed for \protect\modelone{}. A $^{\dagger}$ means that for \modelone{} this parameter is provided as a input feature to the model.}
    \centering
    
    \begin{tblr}{
  hline{1-2} = {2-5}{},
  hline{8} = {2-5}{},
  cell{-}{2} = {l},
  cell{-}{3} = {c},
  cell{-}{4} = {l},
}
        & \textbf{Parameter} & \textbf{Prior} $\in$[\textbf{Value/Range}] & \textbf{Description} \\
         \parbox[t]{2mm}{\multirow{5}{*}{\rotatebox[origin=c]{90}{\textit{Free parameters}}}} 
        & $z^\dagger$ & Uniform $\in$ [$0,~14$] & Redshift \\
        & $\log_{10} \rm M_\star/M_\odot$ & $\log_{10}$ $\in$ [$4,~12$] & Logarithm of formed stellar mass\\

         & $\rm A_V$ & $\log_{10}$ $\in$ [$-3,~0.7$] & V-band dust attenuation (AB mag) \\
         & Z$_\star$ & $\log_{10}$  $\in$ [$-4,~-1.3$] & Stellar metallicity where $\log_{10}\rm Z_\odot = -1.7$\\
         & sSFR & $\log_{10} \in$ [$-12,~-7$] & Specific star formation rate ($\rm yr^{-1}$) \\
         & $t_{25/50/75}$ & Uniform $\in$ [$0, ~1$] & Normalized fraction of lookback time at which 25/50/75\% of the total mass had formed. \\
         
         \parbox[t]{2mm}{\multirow{5}{*}{\rotatebox[origin=c]{90}{\textit{Fixed parameters}}}} & $\log U$ & $-2$ & Ionization parameter \\
         & $f_{\rm esc}$ & $0.0$ & Ionizing photon escape fraction \\
         & $f_{\rm esc, \ Ly\alpha}$ & $0.1$ & Escape fraction of Lyman-$\alpha$ photons \\
         & $n$ & $-0.7$ & Power-law slope of \cite{calzetti2000dust} attenuation \\
         & $\rm T_{dust}$  & $40$ K & Dust Temperature (Greybody)\\
         & $\beta$  & $1.5$ & Dust Emissivity (Greybody) \\
         & $\alpha$ & $1$ & Dirichlet concentration parameter for GP-SFH models \\

    \end{tblr}

    \label{tab:priors}
\end{table*}

The size of the training dataset required for a robust posterior estimator depends on the dimensionality of the inference problem and the neural network architecture. Higher dimensional models will need significantly larger training datasets to provide adequate sampling of the parameter volume. For models in the literature with comparable dimensionality these can range from $10^{4}$ to a few $\times 10^6$; \cite{choustikovInferringIonizingPhoton2025} use only 13,800 high-fidelity observations (forward modelled photometry from the SPHINX simulation), whilst \cite{iglesias-navarroSimulationbasedInferenceGalaxy2025} used $10^6$ and \cite{wangMissingDataAmortized2024} used $\approx 2\times10^6$. To ensure efficient coverage of the parameter space, we use Latin Hypercube (LHC) sampling \citep{mckay2000comparison}. For our 8-parameter problem, we generate models with both $10^5$ and $10^6$ samples, which we utilize in later sections for different purposes. 

For each parameter sample, we use \synthesizer{} to compute the full SED and derive noiseless observer-frame photometry in a wide range of HST (ACS, WFC3/IR) and JWST (NIRCam, MIRI) filters. The noise is not added at this stage; instead, it is applied dynamically during the feature engineering phase, allowing the same base library to be used for applications to surveys with different depths. The entire generation process can be parallelized using MPI over multiple HPC nodes and the outputs are stored in a series of compressed HDF5 files, in a standardised and self-documented format. By simply importing an MPI python package and running with \texttt{mpirun} this can easily be distributed over HPC resources, with no additional user configuration required. 


\section{Training SBI Models}
\label{sec:training}

This section details the process of training our SBI model, \modelone{}, to infer galaxy physical parameters (e.g., stellar mass, SFH, dust attenuation) from broadband photometric observations. The process involves two main stages: first, constructing the input data arrays, and second, optimizing the neural network architecture and training procedure.

\subsection{Feature Engineering}
\label{sec:features}
Starting with the comprehensive library of simulated photometry generated in the previous section, we perform several feature engineering steps to construct the final parameter ($\theta)$ and observation ($x$) arrays for training. This process is crucial for tailoring the general simulation library to a specific application and for ensuring the numerical stability of the training process. Key steps include selecting the specific photometric filters or wavelength range and resolution that match our observational dataset, transforming the data to a more suitable unit system and optionally applying a realistic noise model to the simulated fluxes. We also include measurements of photometric uncertainty in each filter as part of the observation array for each sample to allow the model to condition posterior estimates on the known uncertainty in the photometry. 

For the application in this paper, we select the 14 photometric filters from HST/ACS and JWST/NIRCam that provide coverage over the full GOODS-South field, as detailed in \autoref{sec:data}. We then apply a custom, empirically derived noise model to the simulated fluxes to match the noise characteristics of the JADES survey data, a process we describe in \autoref{sec:noise_modelling}.

For numerical stability and to handle non-detections gracefully, we convert all photometric fluxes into \textit{asinh} magnitudes \citep{lupton2004preparing}, which have useful properties for inference. This unit system behaves logarithmically for bright sources but is linear for faint sources, avoiding issues with negative or zero fluxes. For a flux $f$ the scaling to an \textit{asinh} magnitude $m$ is as follows
\begin{equation}
    m =-2.5 \log_{10}(e)\left[{\rm asinh}\left(\frac{f}{2f_{b}}\right)+{\rm ln}\left(\frac{f_{b}}{3631 ~{\rm Jy}}\right) \right]
    \label{eqn:m_asinh}
\end{equation}

\noindent where  $f_b$ is a softening parameter, which we set to the filter-dependent 5$\sigma$ flux limit of our observational data. The corresponding magnitude error is

\begin{equation}
    \delta m = \frac{2.5\log_{10}(e)\times f}{\sqrt{f^2 + (2f_b)^2}}
    \label{eqn:m_asinh_err}
\end{equation}
We set the scaling parameter $f_b$ to the 5$\sigma$ detection level for each filter, which we discuss further in \autoref{sec:noise_modelling}. The relationship between the AB and \textit{asinh} magnitudes, and the role of the softening parameter $f_b$ is illustrated in the upper panel of \autoref{fig:noise-model}.

The final input observation vector for our fiducial model, \modelone{}, consists of 29 features: the \textit{asinh} magnitude and its associated error in each of the 14 filters, plus the galaxy's redshift, which we treat as a known quantity. The target parameter vector consists of 11 parameters. This includes the 7 primary parameters which fully parametrize the model; stellar mass ($\log_{10}\rm M_{\star}$), metallicity ($\log_{10}\rm Z_\star$), dust attenuation ($\rm A_V$), four SFH parameters ($\rm SFR, t_{25}, t_{50}, t_{75}$, but also four derived parameters: SFR averaged over 10 Myr ($\rm SFR_{10}$), surviving stellar mass ($\log_{10} \rm M_\star^{surv}$), mass-weighted age ($t_{M_\star}$), and UV slope ($\beta$). SBI lets us directly infer these derived parameters alongside the primary parameters which control our model. We use the same broad, uninformative priors for training the SBI model as were used to generate the initial simulations (see \autoref{tab:priors}.) For the derived parameters we simply use broad uniform priors between the minimum and maximum values present in the training dataset.

\subsection{Noise Modelling}
\label{sec:noise_modelling}

The output of our \synthesizer{} SED modelling is noiseless photometric fluxes we will call $x$. For a full-forward model which matches our observed data $x_{\rm obs}$, we need a process to construct $x'$, noisy photometry which reproduces the noise characteristics of $x_{\rm obs}$.

The simplest approach, treating each filter as independent and given a measurement of the average depth in each filter $\sigma_{f}$, would be to say $x_{f}' = \mathcal{N}(x_{f}, \sigma_{f})$ for each filter $f$, where $\mathcal{N}$ is the Normal distribution. This is an approximation, as in reality $\sigma_f$ will be dependent on $x_{f}$, and is likely to vary across the survey. It will also in some cases correlate with the depth in other filters, e.g. in surveys with a central deep region plus peripheral shallower imaging, e.g. the Hubble Ultra Deep Field (HUDF). 

If the model is provided the photometric uncertainty estimate, in order to condition the posteriors to account for photometric noise, then we note that strictly this would not be $\sigma_f$, but rather $\sigma_f'$, the prediction of the uncertainty given the scattered flux $x'$. For real observations we can not know the true kernel which scattered the photometry, and must estimate the uncertainty dependent on the noisy observation.

For \modelone, we construct our per-filter noise modelling to match the noise properties of the observational dataset we introduce in \autoref{sec:data}. Given the per-band flux and local flux uncertainty estimates produced by the \galfind{} catalogue creation (outlined in \autoref{sec:data}) process for all sources in the catalogue, we first calculate the median 5$\sigma$ depth, which we use for the \textit{asinh} magnitude parameter $f_{b}$. Then we convert flux and flux error to \textit{asinh} magnitudes following \autoref{eqn:m_asinh} and \autoref{eqn:m_asinh_err}, and bin them into 20 uniformly distributed magnitude bins. For each bin we compute the median and standard deviation in the magnitude error, which we use to define two linear interpolators $f_m$ and $f_\sigma$ to predict $x_{m}$ and $\sigma_{m}$ given an input \textit{asinh} magnitude $m$. 

For an input magnitude $m$ the process to generate the scattered magnitude and magnitude error is as follows:
\begin{enumerate}
    \item Predict $x_{m}=f_m(m)$  and $\sigma_{m}=f_\sigma(m)$ given $m$
     
     \item Draw noise kernel $\phi = \mathcal{N}(x_m, \sigma_m)_{[0,\infty]}$
     \item Draw noisy magnitude $m' = \mathcal{N}(m, \phi)$
    \item Predict magnitude error $\sigma' = f_\sigma(m')$
\end{enumerate}

\begin{figure}
    \centering
    \includegraphics[width=\columnwidth]{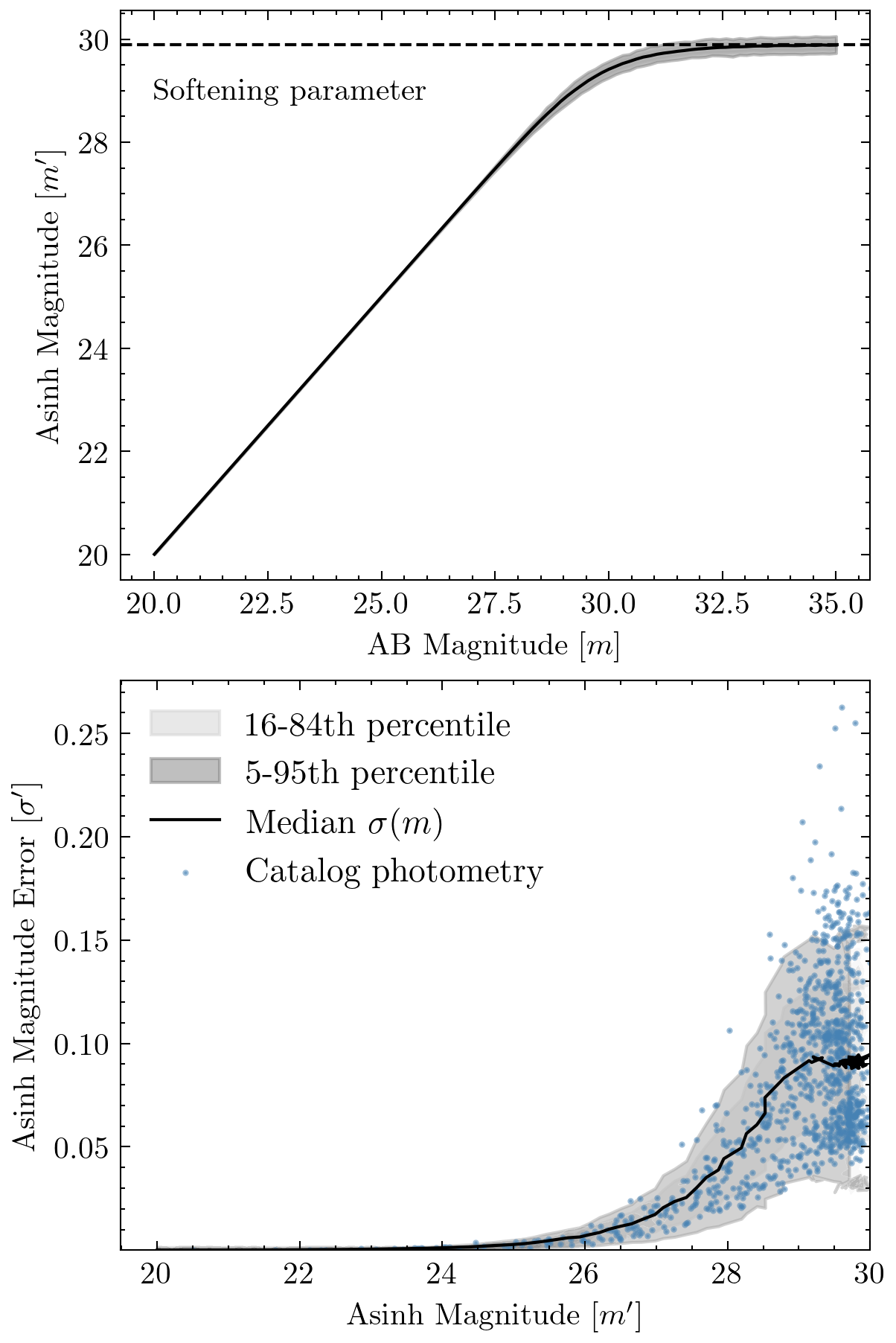}
    \caption{An illustration of the empirical noise model for the F444W filter. The upper panel shows the relationship between the noiseless input magnitude ($m$) and the scattered, noisy magnitude ($m'$). The lower panel shows the relationship between the input magnitude and the assigned photometric uncertainty ($\sigma'$). The blue points show true flux and uncertainty measurements from the JADES photometric catalogue.}
    \label{fig:noise-model}
\end{figure}

This process realistically reproduces the noise properties of the observational data, as illustrated for the F444W filter in  \autoref{fig:noise-model}, which shows the median and 1/2$\sigma$ regime of our noise model relative to a random sampling of the true photometric uncertainties drawn from the JADES photometric catalogue. 

\subsection{Model Architecture and Hyperparameters}

We use \ltuili{} as the backend for training our SBI model, which provides a unified interface to established libraries such as \textsc{sbi} \citep{tejero2020sbi} and \textsc{lampe}\footnote{\url{https://github.com/probabilists/lampe}}. As listed in \autoref{tab:hyperparam}, a variety of neural network architectures are supported by these backends, but for \modelone{} we test two density estimators: the flexible \textit{neural spline flow} \citep[NSF, ][]{durkanNeuralSplineFlows2019} and the \textit{mixture density network} \citep[MDN, ][]{bishop1994mixture}.

These density estimators can be sensitive to the hyper-parameters controlling the depth and width of the neural network. For a NSF this is parametrized as \texttt{hidden\_features} and \texttt{num\_transforms}, where the number of hidden features control the width of the hidden layers in the neural networks which make up the spline transformations and the number of transforms controls the number of invertible transforms composed into the flow. 

To find the optimal configuration for our problem, we use the \textsc{Optuna} framework \citep{akiba2019optuna} in order to determine the best hyperparameters for our model. We performed approximately 3000 parallel trials, exploring the parameter space listed in \autoref{tab:hyperparam}. Each trial utilized the \modelone{} realization with $10^5$ samples for computational efficiency and we used \textsc{Optuna}'s \texttt{MedianPruner} to terminate unpromising trials early, thereby accelerating the search. The objective was to maximize the validation log-probability.

\begin{table}

\caption{An overview of the relevant model architectures available in \ltuili{}, and the hyperparameter ranges and optimal values used in our \textsc{Optuna} optimization.}

\resizebox{1 \columnwidth}{!}{
\begin{tabular}{lll}
\multicolumn{3}{c}{\textbf{Backends and Network Architectures}}\\\hline       
\multicolumn{3}{l}{\textsc{sbi}: Mixture density network \citep[MDN, ][]{bishop1994mixture};} \\\multicolumn{3}{l}{\ \ \ \ \ \ \ Masked autoregressive flow \citep[MAF, ][]{papamakariosMaskedAutoregressiveFlow2018};} \\ 
\multicolumn{3}{l}{\ \ \ \ \ \ \ Neural spline flow \citep[NSF, ][]{durkanNeuralSplineFlows2019};} \\
\multicolumn{3}{l}{\ \ \ \ \ \ \ Masked Autoencoder for Distribution Estimation}\\ 
\multicolumn{3}{l}{\ \ \ \ \ \ \ \ \ \  \ \citep[MADE, ][]{germain2015made}}  \\
\multicolumn{3}{l}{\textsc{lampe}: MDN; MAF; NSF;}\\
\multicolumn{3}{l}{\ \ \ \ \ \ \ Continuous normalizing flow \citep[CNF, ][]{chen2018neural,grathwohl2018ffjord}} \\
\multicolumn{3}{c}{\textbf{Hyperparameters}}\\ \hline
\textit{Free Parameter Name} & \textit{Range} & \textit{Optimal Value} \\
NDE & NSF $\lor$ MDN & NSF \\
Sampler & Adam $\lor$ AdamW & Adam \\
Learning Rate &  [$5\times 10^{-2}~-~10^{-6}$] & 0.0007\\
Number of transforms (NSF) & [$3 - 20$] & 14 \\
Hidden features (NSF) & [$10 - 100$] & 30 \\
Hidden features (MDN) & [$10 - 200$] & - \\
Number of components (MDN) & [$10 - 600$] & - \\
Training batch size & [$32 - 256$] & 79 \\
Stop after epochs & [$10 - 60$] & 57 \\
Max Gradient & [$0.1-10]$ & 6.66\\
\hline
\textit{Fixed Parameter Name} && \textit{Value} \\
Validation Fraction &-& 10\% \\
Train-Test Fraction &-& 10\% \\
\hline  
\end{tabular}}
\label{tab:hyperparam}
\end{table}

\begin{figure*}
    \centering
    \includegraphics[width=\linewidth]{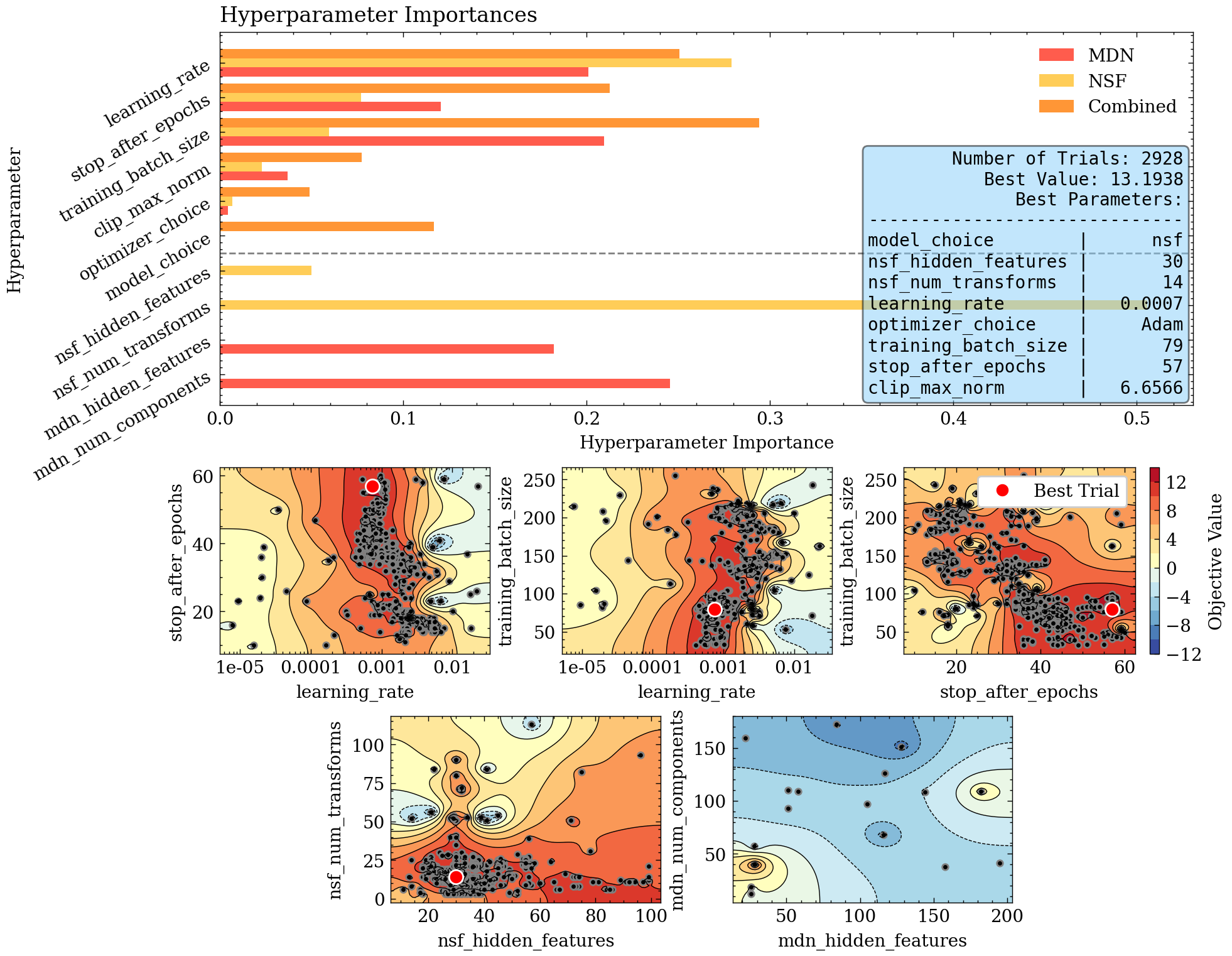}
    \caption{\modelone{} NPE hyperparameter importance and logarithmic P($\theta|x)$ contours for key hyperparameters as determined by Optuna testing \protect\citep{akiba2019optuna}. The upper panel shows the importance of each hyperparameter to the objective optimization for the overall trial, as well as individually for the mixture density network (MDN) and neural spline flow (NSF) network architectures we tested. The lower panels show objective contours for hyperparameter pairs for the learning rate, training batch size, early stopping criteria and network specific parameters controlling the width and depth of the neural networks. Based on $\sim$3000 trials, the optimal model utilized a NSF architecture with hyperparameters given in the inset table.}
    \label{fig:optuna}
\end{figure*}

The results of this optimization are shown in \autoref{fig:optuna}. The upper panel shows parameter importance for both the full set of trials and for the NSF and MDN network architectures alone. The NSF architecture was consistently preferred over the MDN in our trials, achieving higher log-probabilities. The lower panel shows log probability contours for a subset of the optimization parameters; the learning rate, training batch size, early stopping criteria, and the number of hidden features and transforms/components for the NSF and MDN. 

The optimization metric had a significant dependence on several hyper-parameters, in particular the number of transforms, the learning rate and training batch size. The optimal trial utilized a NSF with 30 hidden features and 14 transforms, trained with a \cite{kingma2014adam} optimizer, with a training batch size of 79 and a learning rate of 7e-4. Trials with higher numbers of transforms perform more poorly, which may be because the model over-fits to the training dataset, due to the higher number of free parameters in the flow. The best learning rate, which controls the optimizer step size, is well-constrained, with learning rates which are too large or too small jumping over minima or getting stuck in local minima, respectively. Further improvements to the model may be made by utilizing a learning rate scheduler, which decreases the learning rate as the model approaches a minima. The early stopping criteria, which controls the number of epochs without improvements which are trained before training is stopped, is approaching the prior bound, but good results are obtained anywhere from 30 epochs and up, which provide a good compromise between allowing the model to condition well to the data without over-fitting.

\subsection{Model Training and Validation}
\label{sec:validation}

Our final model realization used the optimized hyper-parameters, but trained on a larger model grid of $N=10^6$ observations. This model took $\sim$3 days to train on a desktop CPU (distributed across 20 cores). 10\% of the training dataset was withheld for use during model validation. Model validation consists of testing the accuracy and bias of the model, as discussed in \autoref{sec:benchmark}. Model coverage tests such as SBC and TARP are implemented in \textsc{LtU-ILI} and are run automatically by \synference{}.


As this example model was trained with a noise model matched to observational data from JADES, there are regions of the parameter space (e.g. low mass galaxies at high redshift) where we cannot expect accurate parameter recovery as the noise level dominates over the mock photometry. However an unbiased SBI model should still return the unconstrained prior proposal in the noise-dominated case. When evaluating parameter recovery we limit ourselves to observations within the validation set which have an average signal to noise in the long-wavelength NIRCam wideband filters $\rm SNR > 5$, but we utilize the full test dataset for the TARP and SBC metrics. 

For each parameter we compute the Root Mean Square Error (RMSE) and the co-efficient of determination $R^2$:

\begin{equation*}
    \text{RMSE} = \sqrt{\frac{1}{n} \sum_{i=1}^{n} (x_i - \hat{x}_i)^2}
\end{equation*}

which quantifies the deviation between the true and recovered parameter value. $R^2$ is given by

\begin{equation*}
R^2 = 1 - \frac{\sum_{i=1}^{n} (x_i - \hat{x}_i)^2}{\sum_{i=1}^{n} (x_i - \bar{x})^2}
\end{equation*}

which measures the proportion of the variance in the dependent variable that is predictable from the independent variable, with a value closer to 1 showing a more reliable prediction. In these equations: n is the total number of validation samples. $x_i$ is the i-th true value, $\hat{x_i}$ is the i-th predicted  value, and $\bar{x}$ is the mean of all the true values $x_i$.

\begin{figure*}
    \centering
    \includegraphics[width=0.9\linewidth]{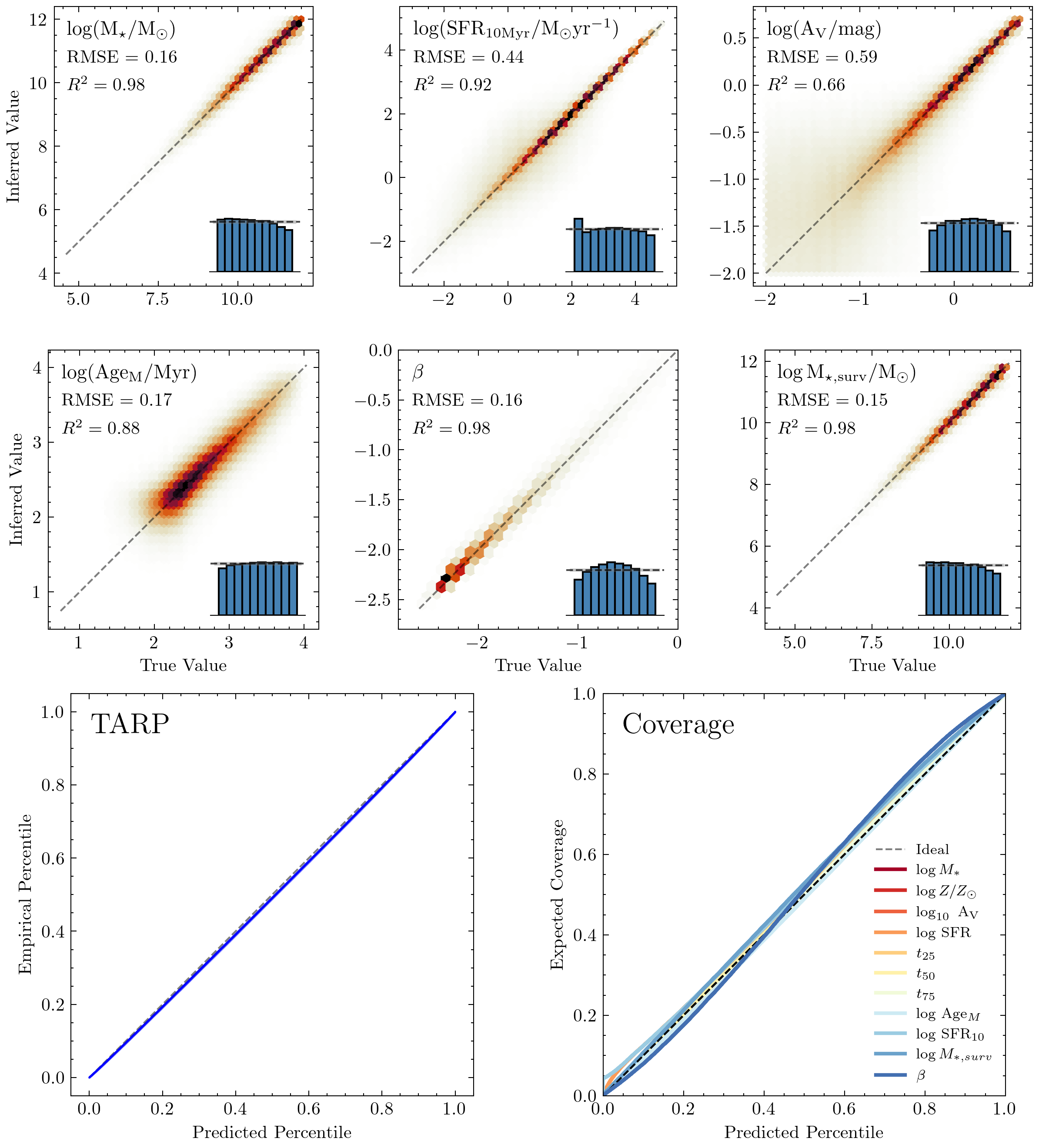}
    \caption{Model performance metrics and calibration tests for \modelone{}. \textbf{Upper panels}: True vs recovered parameter estimates for key model parameters as meausred with the subset of the validation dataset with $\langle \rm SNR \rangle> 5$. Each panel has the posterior rank histogram inset showing the posterior calibration, and the coefficient of determination $R^2$ and root mean square error (RMSE) are provided for each parameter. A rank histogram evaluates calibration by plotting the distribution of the true value's rank relative to sorted samples from the model's posterior predictive distribution; a flat, uniform histogram is ideal, indicating that the true values are statistically indistinguishable from the predicted samples.
    \textbf{Lower right:} Coverage plots for individual parameters showing unbiased estimates. This compares the nominal credible level (the "expected" coverage probability, e.g., 95\%) on the x-axis against the empirical coverage (the actual proportion of true values that fall within that predicted interval) on the y-axis. The 1:1 diagonal line represents perfect calibration, where the model's reported uncertainty (e.g., a 95\% interval) perfectly matches its empirical performance (it contains the true value exactly 95\% of the time). \textbf{Lower left:} Tests of Accuracy with Random Points metric \citep[TARP, ][]{lemos2023sampling} showing the model calibration relative to the ideal 1:1 line. TARP evaluates the average calibration of the model across the entire data distribution. It plots the nominal credible level (x-axis) against the expected coverage probability (y-axis), which is the empirical coverage averaged over all possible data points. }
    \label{fig:model_validation}
\end{figure*}

These results are shown in \autoref{fig:model_validation}. The upper panels show the parameter recovery for a subset of the fitted model parameters, including the total and surviving stellar masses, the star formation rate, dust attenuation, mass-weighted age and UV $\beta{}$ slope (between 1250\AA{} and 3000\AA{}). We plot density hexbins showing the true and recovered values for the full posterior estimates for each parameter. We find generally good parameter recovery for these parameters, with high values for the coefficient of determination $R^2$ suggesting high correlation between true and recovered parameter estimates. Inset into each panel are the rank histograms, which evaluates model calibration
by plotting the distribution of the true value’s rank relative to sorted samples from the model’s posterior predictive distribution; a flat, uniform histogram is ideal, indicating that the true values are statistically indistinguishable from the predicted sample, whereas a histogram which is more bell-shaped means that the learned posterior is too wide. The rank histograms shown are generally flat or slightly bell-shaped indicating that the posterior distributions in \modelone{} could be tighter than is produced by the model.

The individual SFH parameters are more difficult to recover due to significant degeneracy between parameters. In particular with the GP-SFH model, the disconnect between the recent SFR and the historic SFH represented by the $t_{25/50/75}$ parameters means that disparate combinations of these parameters can produce functionally similar star formation histories. To combat this we also infer SFH proxies, such as the 10 Myr averaged SFR and the mass-weighted age, which can break this degeneracy, and show significantly better parameter recovery. 

The lower panels of \autoref{fig:model_validation} showing the overall model `TARP' metric and individual coverage plots for parameters suggests the model is unbiased and produces accurate posterior estimates, as both metrics closely follow the 1:1 line for all parameters. The coverage plots evaluate the nominal credible level (the "expected" coverage probability, e.g., 95\%) on the x-axis against the empirical coverage (the
actual proportion of true values that fall within that predicted interval) on the y-axis. Perfect calibration is represented by the 1:1 diagonal line represents where the model’s reported uncertainty (e.g., a 95\% interval) perfectly matches its empirical performance. \modelone{} shows well-calibrated posteriors for all parameters. 

We also evaluate a range of model benchmarks to evaluate the overall model performance as well as per-parameter metrics, which we show in \autoref{tab:model_performance}. These include the coefficient of determination $R^2$ and the root mean square error. 

\begin{table}
\centering
\caption{Model performance metrics for each parameter, including the coefficient of determination ($R^2$) and Root Mean Square Error (RMSE). The $R^2$ values are calculated across the full validation dataset. The RMSE is also calculated for high-signal-to-noise (SNR) subsets of the validation set (SNR$\geq 5$ and SNR$\geq 20$, as measured in F277W, F356W and F444W). This high-SNR case tests the model's underlying accuracy for parameter recovery, independent of uncertainties from photometric scatter (noise).}
\begin{tabular}{c|cccc}
\textbf{Parameter}                               &$\mathbf{\rm R^2}$ & $\mathbf{\sqrt{|\sigma^2|}}$ & $\mathbf{\sqrt{|\sigma^2_{>5\sigma} |}}$ & $\mathbf{\sqrt{|\sigma^2_{>20\sigma} |}}$\\
\hline
$\log_{10} \rm M_\star/M_\odot$         & 0.99  & 0.75  & 0.21    & 0.18          \\
$\rm \log_{10} Z_\star$                 & 0.96  & 0.72 & 0.43      & 0.37         \\
$\rm \log_{10} A_V$                               & 0.86  & 0.94    & 0.64  & 0.61          \\
$\log_{10} \rm SFR$                     & 0.88  & 1.28  & 0.64    & 0.57         \\
$t_{25}$                                & 0.95  & 0.19  & 0.19    & 0.19          \\
$t_{50}$                                & 0.88  & 0.20 & 0.20      & 0.21          \\
$t_{75}$                                & 0.77  & 0.19  & 0.20    & 0.17          \\
$\rm \log_{10} t_{mwa}$                 & 0.98  & 0.20   & 0.17  & 0.17           \\
$\log_{10} \rm SFR_{10 Myr}$            & 0.90  & 1.15  & 0.55  & 0.48            \\
$\log_{10} \rm M_{\star, surv}/M_\odot$ & 0.99  & 0.73 & 0.20  & 0.63             \\
$\beta$                                 & 0.86  & 1.02  & 0.25  & 0.20            \\
   
\end{tabular}
\label{tab:model_performance}
\end{table}

\subsubsection{Comparison to Nested Sampling}

To obtain reference posterior estimates for observations within our validation dataset we use Bayesian nested sampling implemented through the \textsc{dynesty} package. \textsc{dynesty} provides Bayesian posterior and evidence estimates. We run the dynamic nested sampling algorithm in $\textsc{dynesty}$ using slice sampling to propose new points, implementing multi-ellipsoidal bounding for efficiency. The process starts with 250 live points, and the initial run continues until the estimated improvement in evidence ($\Delta$logZ) drops below 0.1.

We use $\mathcal{L}=-\frac{1}{2}\chi^2$ as our likelihood function, and use the same \synthesizer{} based model to generate mock photometry for each likelihood evaluation. \textsc{dynesty} takes $\sim$90 minutes per fit on a single CPU core, which is computationally prohibitive for the full validation set, so we have chosen a selection of observation vectors from the validation set which cover significantly different regions of the parameter space.

\begin{figure*}
    \centering
    \includegraphics[width=\textwidth]{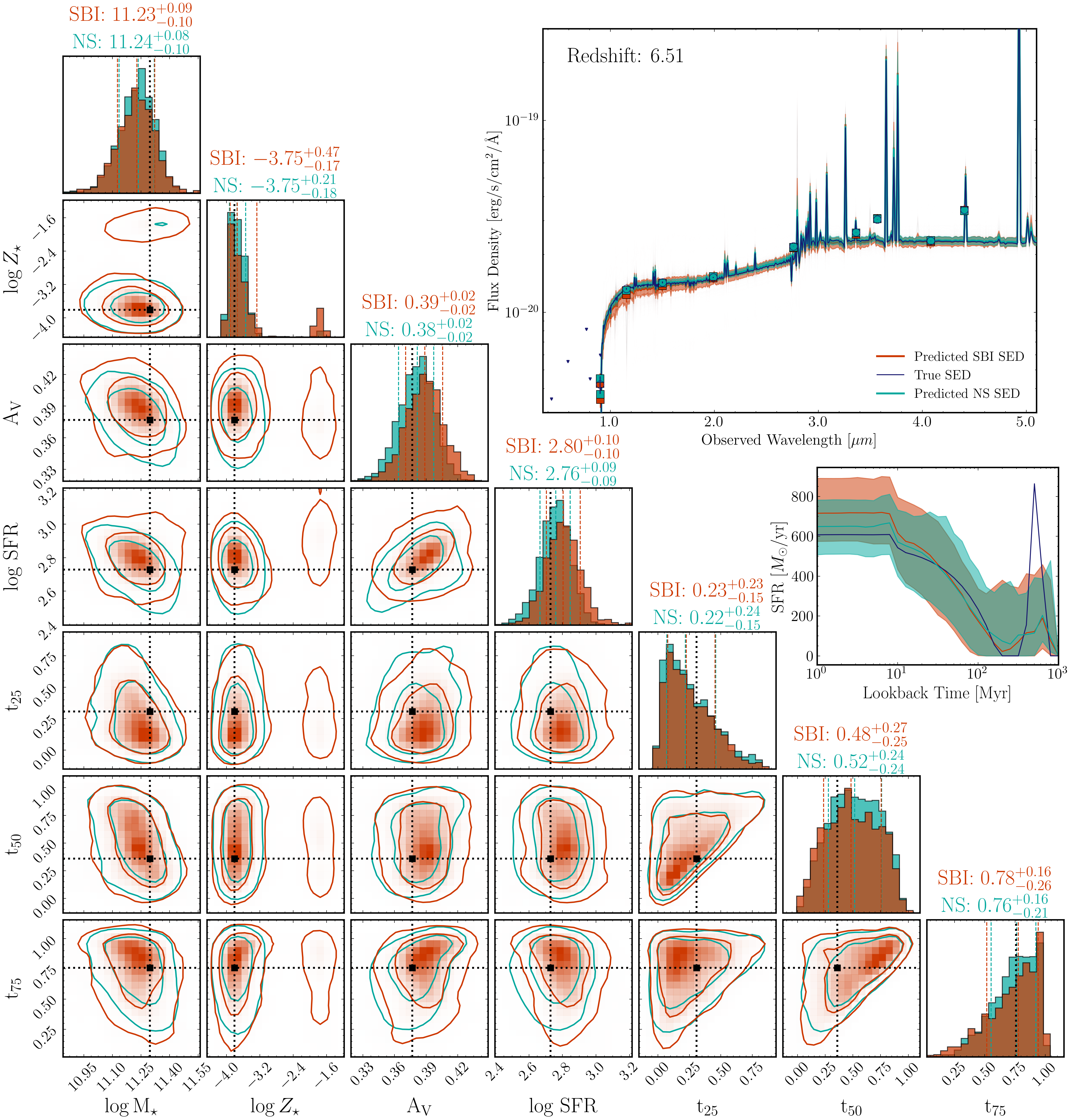}
    \caption{Corner plot for a single mock galaxy comparing multivariate posterior constraints from SBI (in red) to the ground-truth value (black point) as well as reference nested-sampling (labelled as NS) results using \textsc{dynesty} (in blue). Contours show the $1\sigma$ and $2\sigma$ regions. \modelone{} accurately recovers the stellar mass, metallicity, dust attenuation and SFH. In the upper right the true and recovered SED and SFH models are shown showing accurate SFH and SED recovery for this galaxy in both cases.}
    \label{fig:corner}
\end{figure*}

In \autoref{fig:corner} we show a corner plot showing the multivariate posterior distribution for a random example galaxy from the validation set for both \modelone{} and the nested sampling results from \textsc{dynesty}. We also show the ground truth parameter values in black. We observe strong agreement between both the univariate and multivariate posterior distributions for all fitted parameters in this test case. Interestingly, both the  \modelone{} and NS results identify the same slightly bi-model result for metallicity, although they disagree slightly on it's relative probability. We recover the ground-truth values within $1\sigma$ for all parameters. 

We also show the reconstructed SED and SFH for the galaxy which we derive by calculating the predicted SED for each posterior draw, and taking the 16th/50th/84th percentiles of the distribution to find the median and $1\sigma$ confidence interval for both the \modelone{} and NS posterior estimates. The reconstructed SEDs and SFHs both closely reproduce the ground-truth model, and the reconstructed photometry also recovers the input photometry used for the inference, which shows the model has accurately learnt the observation-parameter relationship in this case. 

\subsection{SED Recovery}

\begin{figure*}
    \centering
    \includegraphics[width=0.87\linewidth]{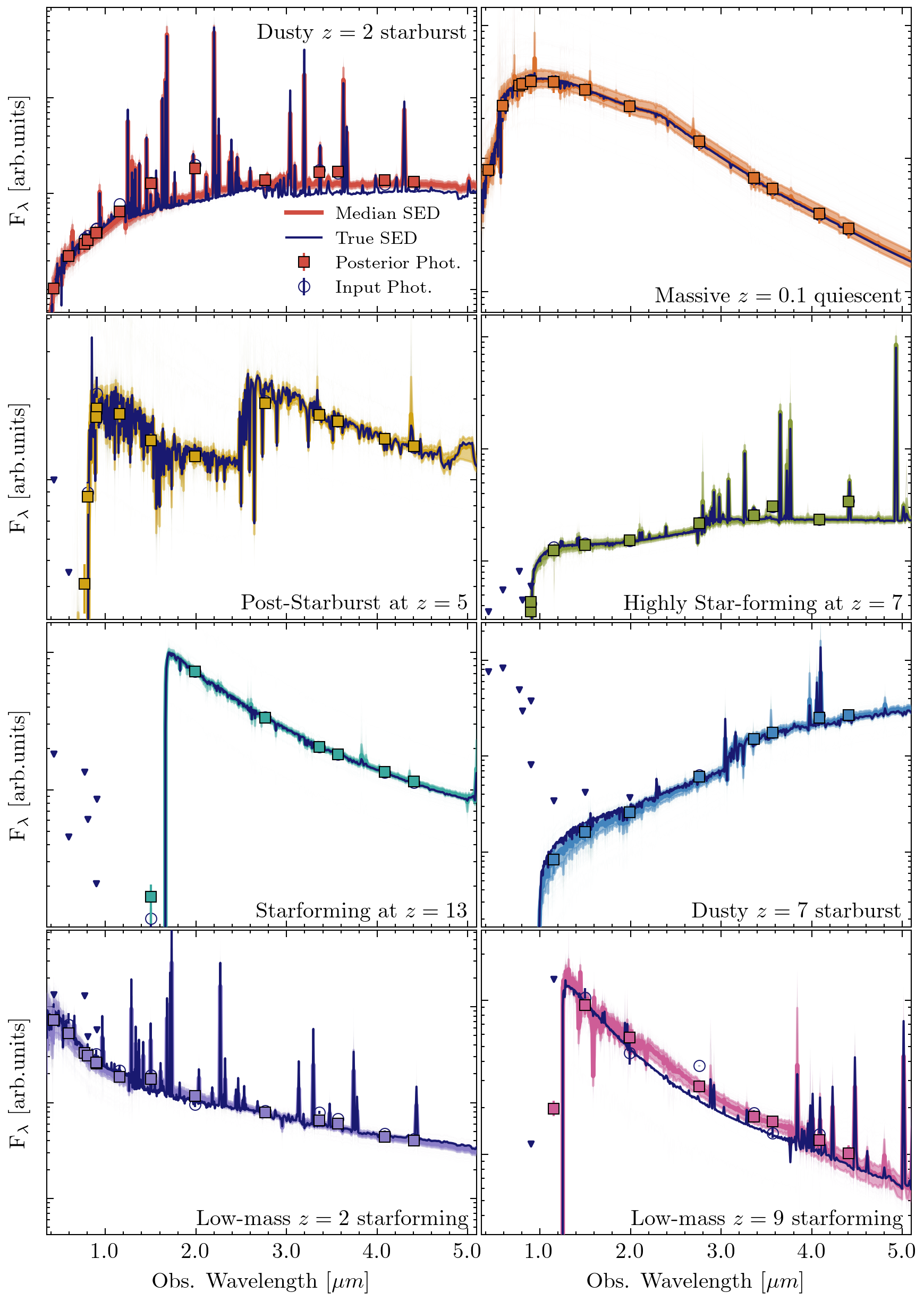}
    \caption{SED recovery with SBI inferred posterior parameters compared to true SED and photometry for 8 distinct regions of model parameter volume. Dark blue lines show the input spectra, and colored lines and shaded regions show the 50th and 16-84th percentile of the reconstructed SED using the posterior samples from \modelone{}. Photometry provided to the model for inference is shown with unfilled circles and downwards triangles, which correspond to measurements with SNR<3 and are plotted as a 3$\sigma$ upper limit for visual clarity given the logarithmic y-axis scaling. The model is able to successfully reproduce the input SEDs across different regions of the parameter hypervolume.}
    \label{fig:SED_recovery}
\end{figure*}

We show the true and recovered SEDs and ACS\_WFC/NIRCam photometry for a range of input galaxy parameters in \autoref{fig:SED_recovery}, which reflect the observed diversity of galaxies. We include galaxies from the local Universe to high-redshift, with SFHs ranging from quiescent to a starburst, and from entirely dust-free to very obscured. The shaded volume for each galaxy show the range of SEDs consistent with the posterior draws for that input photometry, which we compare with the true input SED in dark blue. 

For all eight distinct galaxy models shown the SBI posterior estimates produce SEDs which are close to the true underling galaxy SED, suggesting the model has learnt the observation-parameter relationship across a wide range of input parameters reflecting highly diverse physical galaxy properties.

\section{Application to Observations}
\label{sec:applications}

The purpose of training our SBI model is to perform inference on real observations, where we have no knowledge of the ground truth. Our chosen dataset is a photometric catalogue of galaxies within the GOODS-South field, which is a well-studied region with a large amount of ancillary observations including space and ground-based spectroscopic redshifts. 

\subsection{Data Products and Catalogue}
In this section we detail the photometric and spectroscopic data products used to construct our galaxy catalogue upon which we validate our SBI model.

\label{sec:data}
\subsubsection{Photometric Data Products}
We use the JWST NIRCam imaging of the GOODS-South field taken as part of the JWST Advanced Deep Extragalactic Survey \citep[JADES;][see also \citet{Bunker2023-NIRSpec} and \citet{Hainline2024a}]{Eisenstein2023}. This includes the F090W, F115W, F150W, F200W, F277W, F335M, F356W, F410M, and F444W NIRCam data presented in \citealt{Rieke2023b} (DR1) and additional area from PIDs 1180 (PI: D. Eisenstein), 1286 (PI: L{\"u}tzgendorf), and 1287 (PI: Isaak). HST ACS legacy data for the F435W, F606W, F775W, F814W, and F850LP filters is taken from v2.5 of the Hubble Legacy Fields \citep{Illingworth2016, Whitaker2019-HLF}, for a total of 14 photometric filters. 

\subsubsection{Catalogue Creation}

Our photometric reduction and catalogue creation process uses the \textsc{GALFIND} pipeline\footnote{\url{https://galfind.readthedocs.io}}, which has been previously used in the EPOCHS paper series \citep[see. e.g.][]{adams2024epochs, Conselice_2025, 2024arXiv240410751A, 2025ApJ...978...89H}. For full details of the data reduction and catalogue creation process for this field please see Austin et al. (in prep.). 

In summary, we use a modified version of the official JWST pipeline, with CDRS version 1263, and align the image mosaics to Gaia DR3. For our catalogue creation we use \sextractor{}, and perform detection in an inverse-variance weighted stack of F277W, F356W and F444W. We measure aperture photometry in 0\farcs32 diameter apertures as well as total photometry in elliptical Kron apertures. 

Whilst the typical EPOCHS catalogue creation procedure is optimized for the study of high-redshift galaxies, we wish to validate our SBI methodology over a wider sample of galaxies. Therefore in a deviation from the standard EPOCHS approach we also measure total fluxes for each galaxy in a similar fashion to \cite{weibel2024gsmf} and \cite{2024ApJS..270....7W}. We derive a correction factor from the flux ratio between the 0$\farcs$32 aperture and the Kron aperture in a PSF-matched version of the detection image. In the case that the Kron aperture covers a larger area, and the measured ratio is greater than unity, we correct measured fluxes in all bands by this ratio before any aperture corrections are applied. We then calculate an aperture dependent PSF-correction for each galaxy by placing either the Kron or circular aperture on our F444W PSF model and measuring the enclosed flux. In the case that the Kron aperture extends beyond the $4\arcsec\times 4\arcsec$ extent of our PSF models, we approximate the Kron ellipse as a circle with radius $\tt{KRON\_RADIUS}\times \sqrt{\tt{A\_IMAGE}\times B\_IMAGE}$, and calculate the aperture correction from tabulated enclosed energy tables for NIRCam. This procedure results in flux measurements which account for the total light emitted by each galaxy, but without any distortion of the galaxy colour due to changing Kron apertures in different filters. Our PSF models are created by stacking isolated stars in each mosaic and follow the exact procedures of \cite{Harvey_2025}.  

\subsection{Spectroscopic Data Products}
\label{sec:specz}
We obtain spectroscopic redshift catalogues from previous studies using both space and ground-based spectroscopy, including JWST/NIRSpec \citep{Jakobsen2022}, 3D-HST, MUSE, and numerous smaller surveys including VLT/FORS2 amd VIMOS/VLT\ \citep{Brammer_2012,bacon2010muse}. For NIRSpec observations of the GOODS-South field we use the official JADES DR3 catalogue \citep{2025ApJS..277....4D}. We filter the catalogue by the spectroscopic redshift and its associated quality by:
\begin{equation*}
(\texttt{z\_Spec\_Flag} = A \lor B \lor C) \land (\texttt{z\_Spec} > 0) \;,
\end{equation*}which results in 1333 galaxies with robust spectroscopic redshifts from JWST/NIRSpec. 

For 3DHST we use the v4.1.5 catalogues for both the HST imaging and grism observations \citep{Brammer_2012, Skelton_2014, Momcheva_2016}, which we filter for robust spectroscopic redshifts following
\begin{equation*}
((\texttt{z\_max\_grism} \neq -1) \land (\texttt{use\_zgrism} = \text{True})) \lor (\texttt{z\_spec\_1} \neq -1) \;.
\end{equation*}

For MUSE we obtain the MUSE HUDF DR2 catalogue \citep{2023A&A...670A...4B}. We also obtain the ESO Master Catalogue v3.0, which compiles spectroscopic redshifts from surveys including \cite{2004A&A...428.1043L,vanzella2005spectroscopic,vanzella2006spectroscopic,vanzella2008spectroscopic,vanzella2009spectroscopic,kurk2013gmass}, which target galaxies with $I <24$. We filter the ESO catalogue for robust spectroscopic redshifts based on the quality factor column. 

We perform a hierarchical matching to our master photometric catalogue, by matching each spectroscopic survey to a subset of the catalogue containing only galaxies with photometry (as determined by the flux in the nearest NIRCam filter to the approximate spectroscopic wavelength probed) bright enough to be detected given the sensitivity of the spectroscopic survey. If a galaxy is matched in multiple spectroscopic surveys we prioritize NIRSpec>MUSE>3D-HST>ESO, and keep only the highest priority spectorscopic redshift available to avoid source duplication. We allow a 0\farcs5 offset for MUSE, 3DHST and NIRSpec, and a 1\farcs0 offset for the ESO catalog. We note however that the average offset is only 0\farcs23.

\subsection{Final Catalogue}

\begin{figure}
    \centering
    \includegraphics[width=\columnwidth]{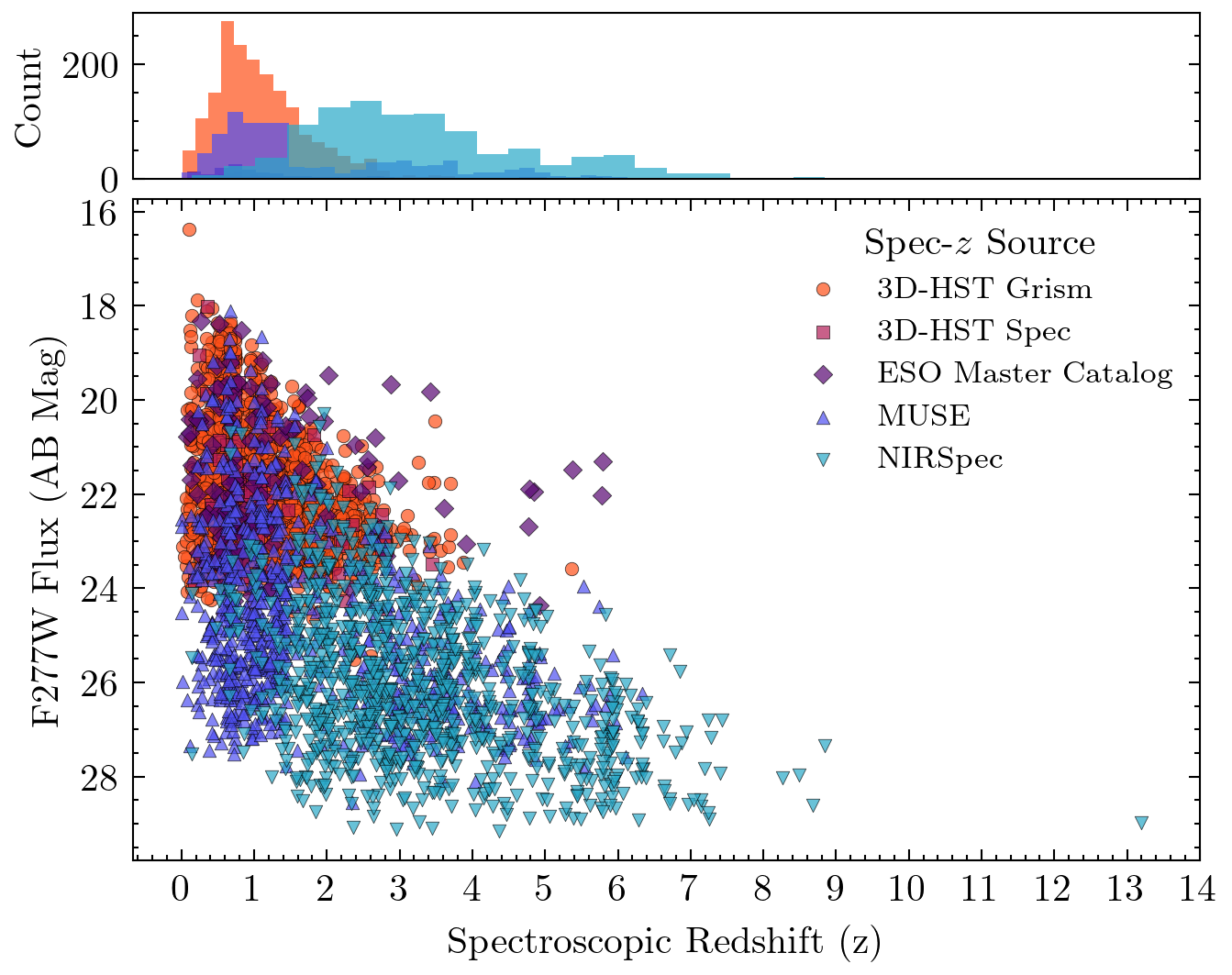}
    \caption{Apparent magnitude in the F277W filter vs redshift for our sample of $\sim$3000 spectroscopically confirmed galaxies in the JADES GOODS-South field. Galaxies are split by the instrument used to determine their spectroscopic redshift, as detailed in \protect\autoref{sec:specz}. The upper panel shows a histogram of the spectroscopic sample as a function of redshift.}
    \label{fig:data}
\end{figure}

Our final spectroscopically-matched catalogue contains 3800 galaxies. We present a magnitude-redshift comparison to demonstrate our sample in \autoref{fig:data}, which includes galaxies from $0 < z < 13.5$.

SBI approaches do not work well with missing data, so for our basic catalogue we require robust flux measurements in the F606W, F775W, F814W, F850LP, F090W, F115W, F150W, F200W, F277W, F335M, F356W, F410M and F444W filters. Approaches in the literature such as \cite{wangSBIFlexibleUltrafast2023} to infer missing observations through Monte Carlo approaches on the training dataset are computationally intensive. This cuts our catalogue down to 3,088 galaxies which have measured photometry in all required filters.

\subsection{Testing model misspecification}

A reasonable check before using any model to perform inference on an observation is that the model is capable of producing mock observations which are similar to the observation itself. This is referred to as "prior predictive checking" \citep{gelman2006prior}. This is particularly important when using an SBI approach, which requires the training data closely matches the observations. In SBI terminology if the model can't reproduce the observations, the model is `mis-specified' to the data, and any inference from the model will be unreliable. 

We carry out prior predictive checking using outlier detection, where we treat our training dataset as a reference distribution against which we test our observations to see whether they are identified as outliers. There are a large number of different approaches to outlier detection, which are suited to different problems. Our outlier detection utilizes majority voting of a suite of different outlier detection algorithms, such that for each observation $\geq50\%$ of the methods must identify the observation as an outlier. 

In \synference{}, prior predictive checks are carried out by default before inference to improve posterior reliability. This is implemented through the \textsc{pyod} Python package \citep{zhao2019pyod}. We utilize a suite of 8 different outlier detection algorithms, including Isolation Forest \citep{liu2008isolation,tony2012isolation}, Local Outlier Factor \citep{he2003discovering}, Deep Isolation Forest \citep{xu2023deep}, Feature Bagging \citep{lazarevic2005feature}, ECOD \citep{li2022ecod}, k-Nearest Neighbors \citep{angiulli2002fast}, Gaussian Mixture Modelling \citep{aggarwal2015data}, and Kernel Density Estimation \citep{latecki2007outlier}. 
Combining different outlier detection approaches averages over the individual biases of each method, and provides a more robust assessment than relying on a single method alone.

We set the outlier contamination at 1\% for all outlier models in the ensemble. Only 0.25\% of our GOODS-South catalogue is identified as a possible outlier by this approach, suggesting our training dataset covers almost the full range of color and luminosity space of our observational catalogue. The galaxies identified as outliers are not a homogenous population, with a wide range in redshift, stellar mass, SFR and $\rm A_V$, based on the \bagpipes{} results. Around 50\% have $\rm A_v> 1$ with \bagpipes{}, so they may be being identified as outliers simply due to our logarithmic dust prior, which favours low dust attenuation. 

\subsection{Galaxy Property Inference}
\label{sec:gal_inference}
We apply our trained amortized NPE model to the photometric catalog to infer parameter posterior estimates.  \synference{} automatically applies the same model scaling and unit conversion to the observations as were applied to the training dataset, which is important for reliable results with SBI. We infer posterior distributions for our spectroscopic catalogue using \modelone{}, which takes ~3 minutes for 3000 galaxies on a typical desktop CPU on a single core (17.5 galaxies/second), which could be further improved when the inference is distributed across multiple threads. 


\begin{figure*}
\centering
    \begin{subfigure}{0.47\textwidth}
    \includegraphics[width=\textwidth]{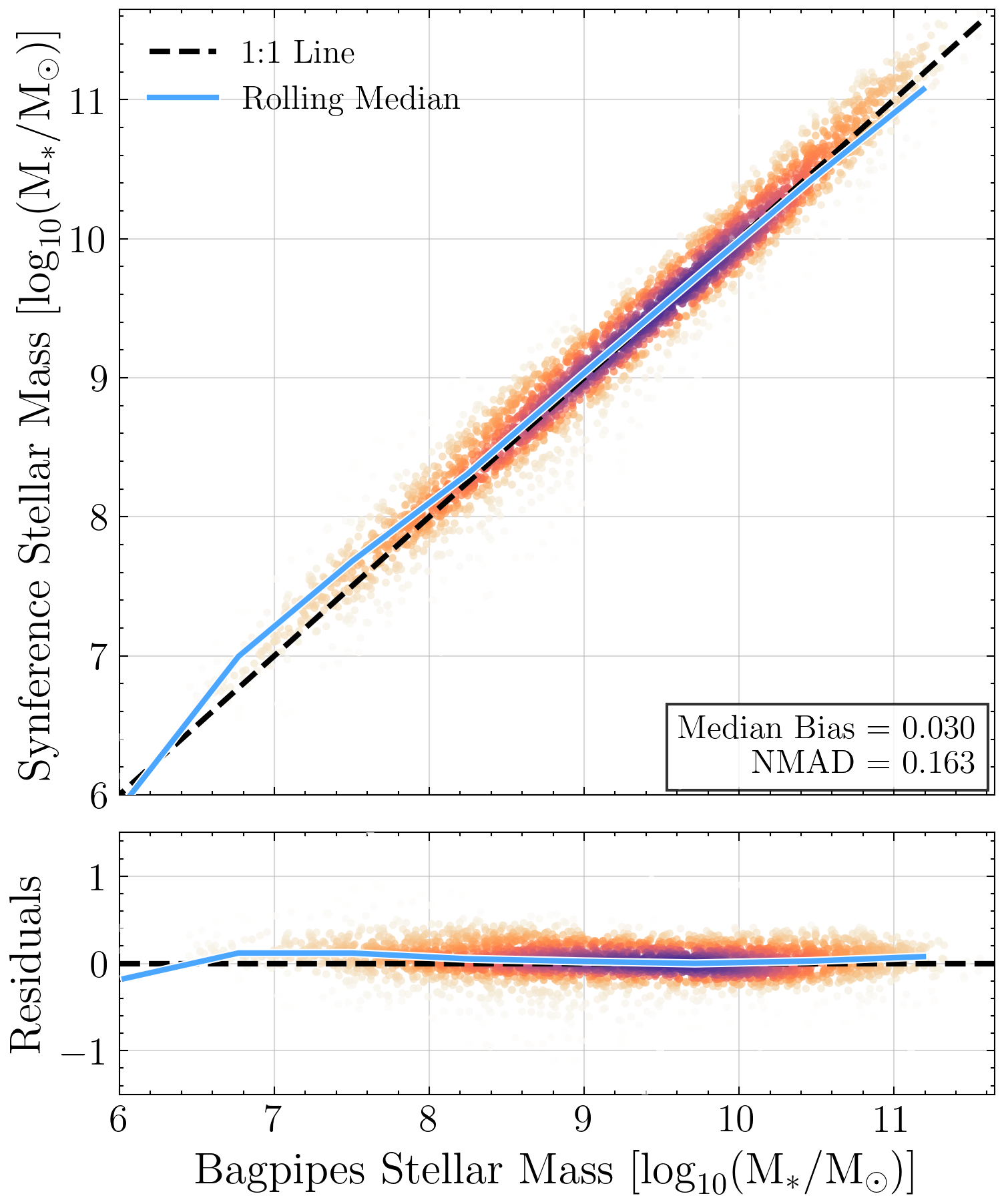}
    \end{subfigure}
   \begin{subfigure}{0.47\textwidth}
    \includegraphics[width=\textwidth]{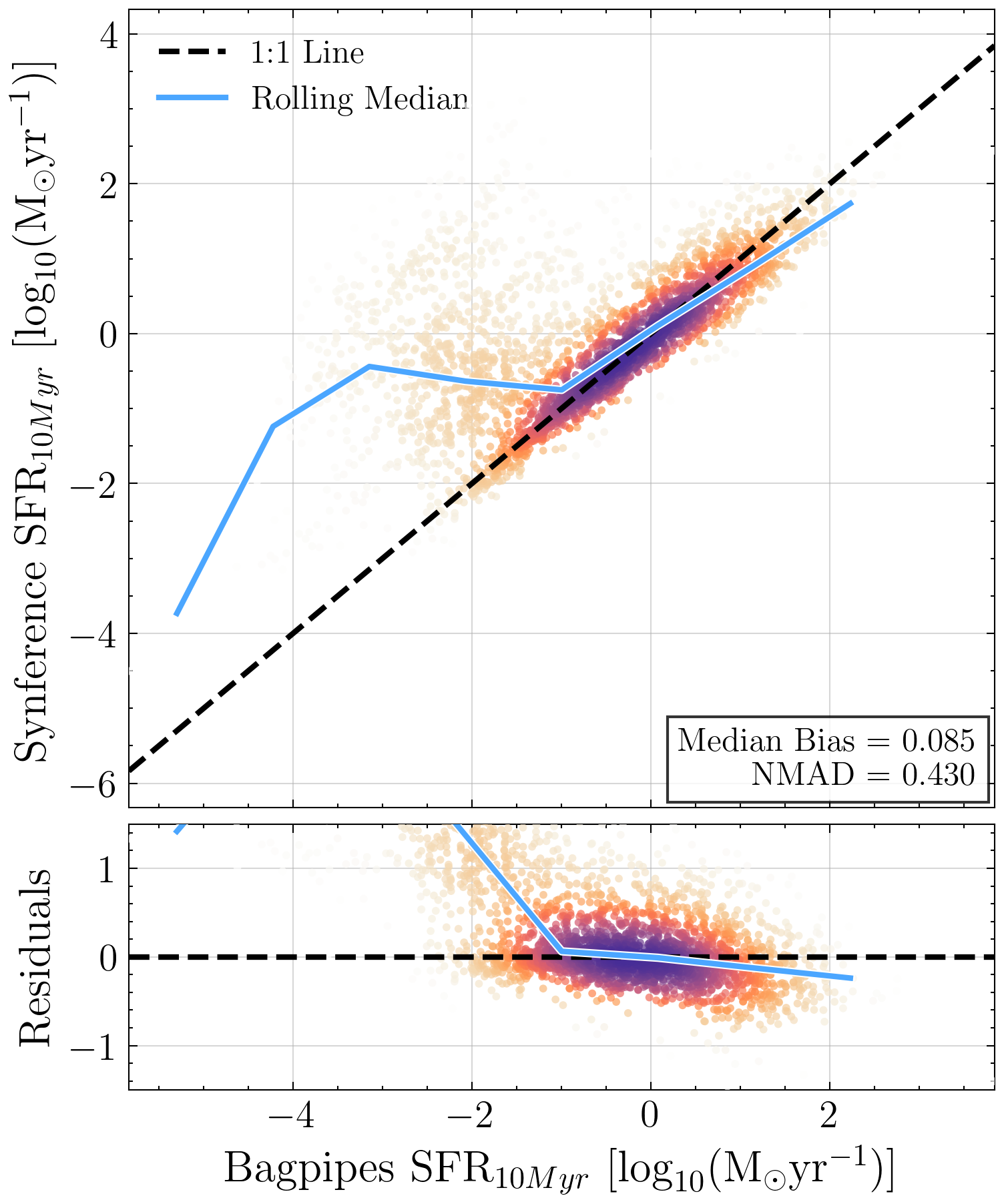}
    \end{subfigure}
    \begin{subfigure}{0.47\textwidth}
    \includegraphics[width=\textwidth]{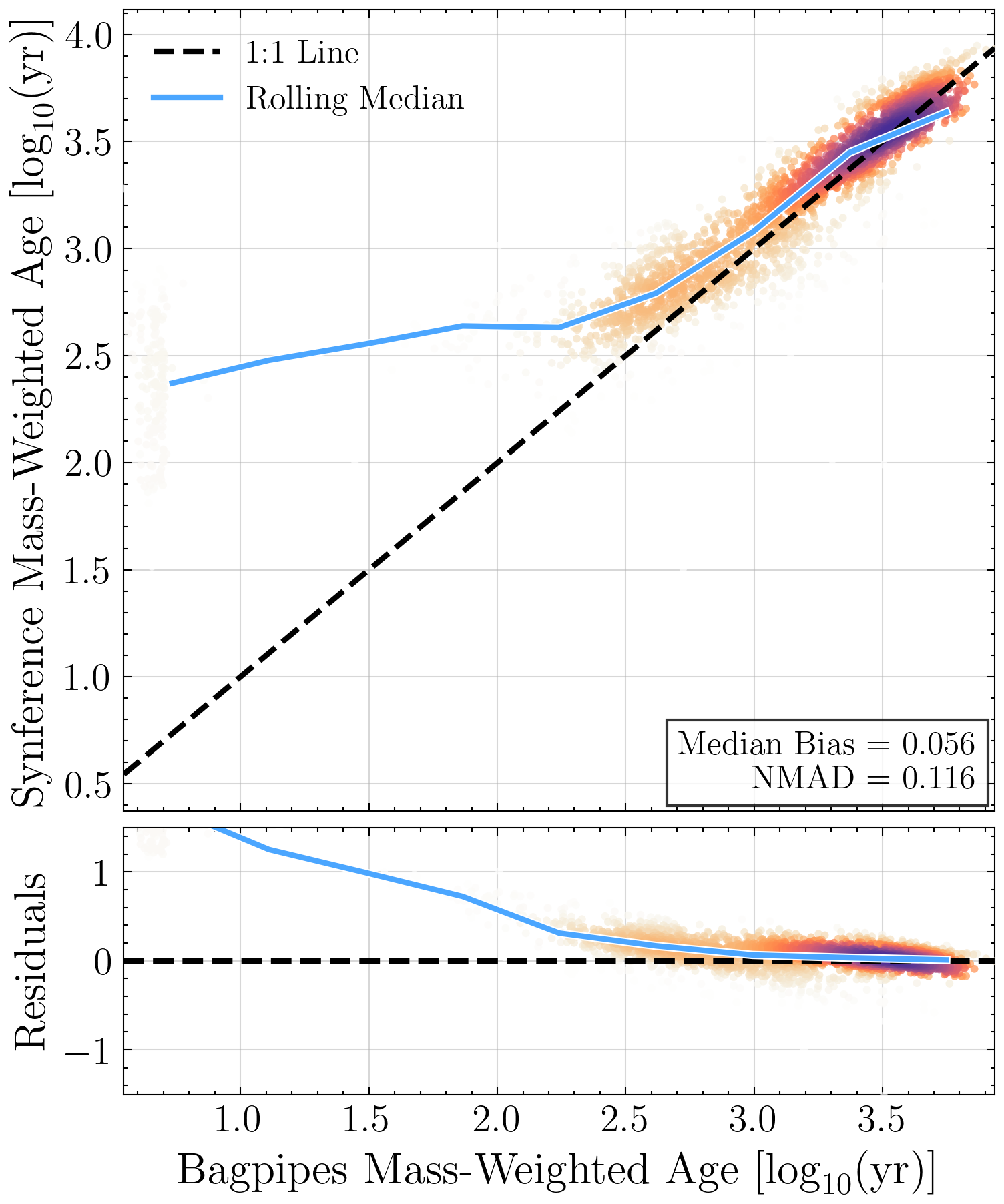}
    \end{subfigure}
    \begin{subfigure}{0.47\textwidth}
    \includegraphics[width=\textwidth]{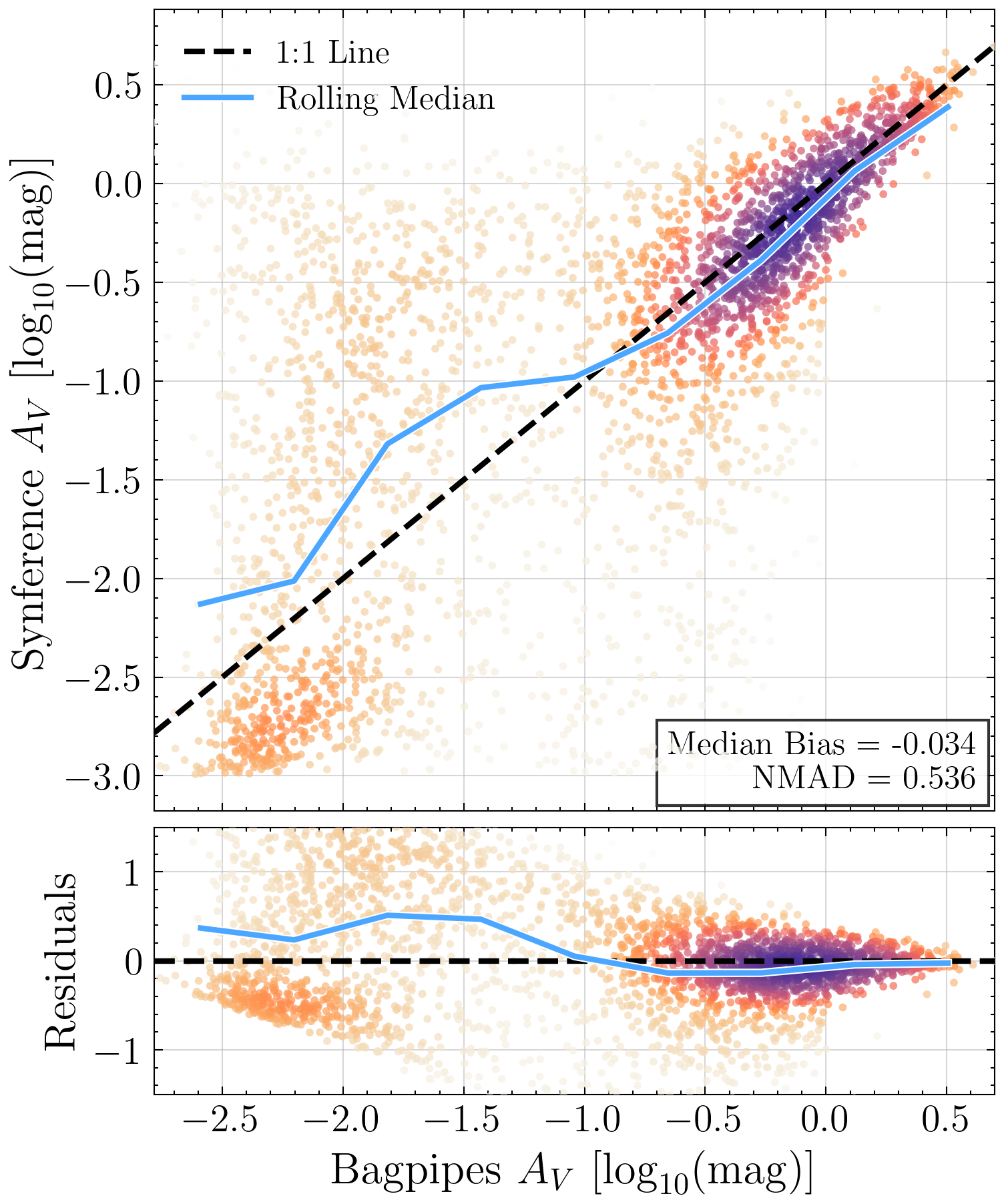}
    \end{subfigure}
    \caption{\bagpipes{} vs. \synference{} physical property inference comparison for the JADES spec-$z$ catalogue, showing stellar mass (upper left), 10 Myr star-formation rate (upper right), mass-weighted age (lower left) and dust attenuation (lower right).  Points are colored by their local density as determined by a Gaussian KDE. The median offset (bias) and the NMAD (Normalized Median Absolute Deviation), a robust measure of the scatter, are inset into the panel for each parameter. Overall we find good agreement in recovered parameters with \bagpipes{}, although there are some systematic deviations in certain regions of the parameter space. The inferred stellar mass shows little bias and relatively low scatter, whereas SFRs and stellar ages are typically systematically higher when using \synference{}. Dust attenuation shows good agreement when there is significant dust present, but shows significant scatter for low dust attenuation, where the distribution skews towards the median of the prior.  The deviation from the 1:1 line for low ages is due to \bagpipes{} inferring a population of galaxies where all star-formation has occurred in a single burst immediately prior to observation, whilist \synference{} prefers a slightly longer SFH.}
    \label{fig:bagpipes_vs_synference}
\end{figure*}


As a point of comparison we also fit the full galaxy sample with \bagpipes{}, sampling using \textsc{MultiNest} \citep{feroz2009multinest}. We use a modified version of \bagpipes{} which can implement the same underlying SPS grid, priors and SFH model as \modelone{} in order to obtain as representative a comparison as possible. The total time for this inference with \bagpipes{} is $\sim$80 CPU-hours, which is 1700$\times$ slower than the SBI model. We estimate parameter values and 1$\sigma$ uncertainties from the 16/50/84th percentiles of the posterior distributions for each parameter.

\autoref{fig:bagpipes_vs_synference} shows a comparison of the recovered stellar mass, star formation rate, dust attenuation and mass-weighted age between the \bagpipes{} and \modelone{} results. The stellar mass comparison for both models is the surviving stellar mass, which excludes stellar remnants and mass loss. These plots show the majority of the galaxies within the density contours, and only galaxies <20\% of the peak density are shown with individual points and errorbars. We also show a rolling median in blue highlighting the overall trend. 

In the upper left-hand panel the stellar mass estimates with \modelone{} and \bagpipes{} show excellent agreement, with a median offset of 0.03~dex. There are a few extreme outliers, which are primarily galaxies with both high-mass quiescent (with strong Balmer breaks) and low-mass (highly star-forming) solutions, on which \bagpipes{} and \synference{} disagree.

The SFR estimates also show reasonable agreements in most cases, but have more systematic offsets particularly at very low star-formation rates, where \synference{} is finding systematically higher star-formation rates. The overall median offset is 0.086~dex. This in part likely due to a slight prior difference between \synference{} and \bagpipes{}. For \synference{} each galaxy has an individual flat sSFR prior of $-12 \leq \log {\rm sSFR/yr} < -7$, whereas this has to be set globally with \bagpipes{}. This leads to a higher lower limit on sSFR for massive galaxies using \synference{}, as there is little observational difference between different sSFR values within the quiescent regime.

The lower left panel shows the mass-weighted age comparison between \synference{} and \bagpipes{} which also shows reasonable agreement, particularly for higher stellar ages. The median offset is 0.057~dex. There are a group of outliers in the \bagpipes{} results which are fitted with unphysical extremely young SFHs, where all mass formed in a very recent burst. These galaxies fall into the region of the Dense Basis SFH parameter space where the SFH is controlled only by the recent SFR parameter, rather than the lookback time quantiles. This may be causing an issue with the nested sampling causing the fitting to perform poorly and not find the optimal solution.

Finally for the dust attenuation, $\rm A_V$, we also see good agreement, albeit with a wider scatter. For very low dust attenuation values we see a large amount of scatter, which is primarily due to the logarithmic scale, as there is very little difference observationally between $\rm A_V = 10^{-2}$~mag and $\rm A_V = 10^{-3} mag$. The median offset is -0.033~dex.

Overall whilst there are some systematic differences between the \bagpipes{} and \synference{} results, they are no worse than other comparisons between different SED fitting codes, which often show different solutions when fitting the same galaxies \citep[e.g.][]{2025ApJ...978...89H}.

\section{Discussion}

In this section we briefly discuss the overall model performance, possible applications, limitations, and make comparisons with existing SED fitting tools. 

\label{sec:discussion}

\subsection{Posterior Accuracy}
\label{sec:performance}

The performance metrics on the validation data, nested sampling comparison and \bagpipes{} comparison show that the fiducial model performs well, and can accurately recover parameters such as stellar mass, star-formation rate, dust attenuation and stellar age to a reasonable accuracy. However, as demonstrated by the average RMSE at different SNR limits in \autoref{tab:model_performance}, some parameters are still affected by significant scatter even at high SNR, which is independent of the error introduced by the photometric scatter. This is particularly true for parameters which are significantly degenerate with one another, such as the 4-parameter Dense Basis SFH. Whilst the recent SFR and overall age of the stellar population can be constrained when reconstructing the full SFH, \modelone{} sometimes produces uncertain results for the overall SFH, where the lookback time quantiles ($t_{25/50/75}$) posteriors are often prior-driven. We note that these degeneracies are not specific to this model but are implicit simply due to the information available from photometry alone. They are not induced by the neural density estimator, and are also observed in the nested sampling and \bagpipes{} results (as can be seen in the reconstructed SFH in \autoref{fig:corner}, and would be observed in any SED fitting code when fitting the same observations.

\subsection{Speed of Inference}

Here we evaluate the speed of amortized inference with SBI compared to traditional Bayesian SED fitting as well as maximum likelihood modelling/optimization algorithms, i.e. for photometric redshifts with \eazy{}.

\begin{figure}
    \centering
    \includegraphics[width=\columnwidth]{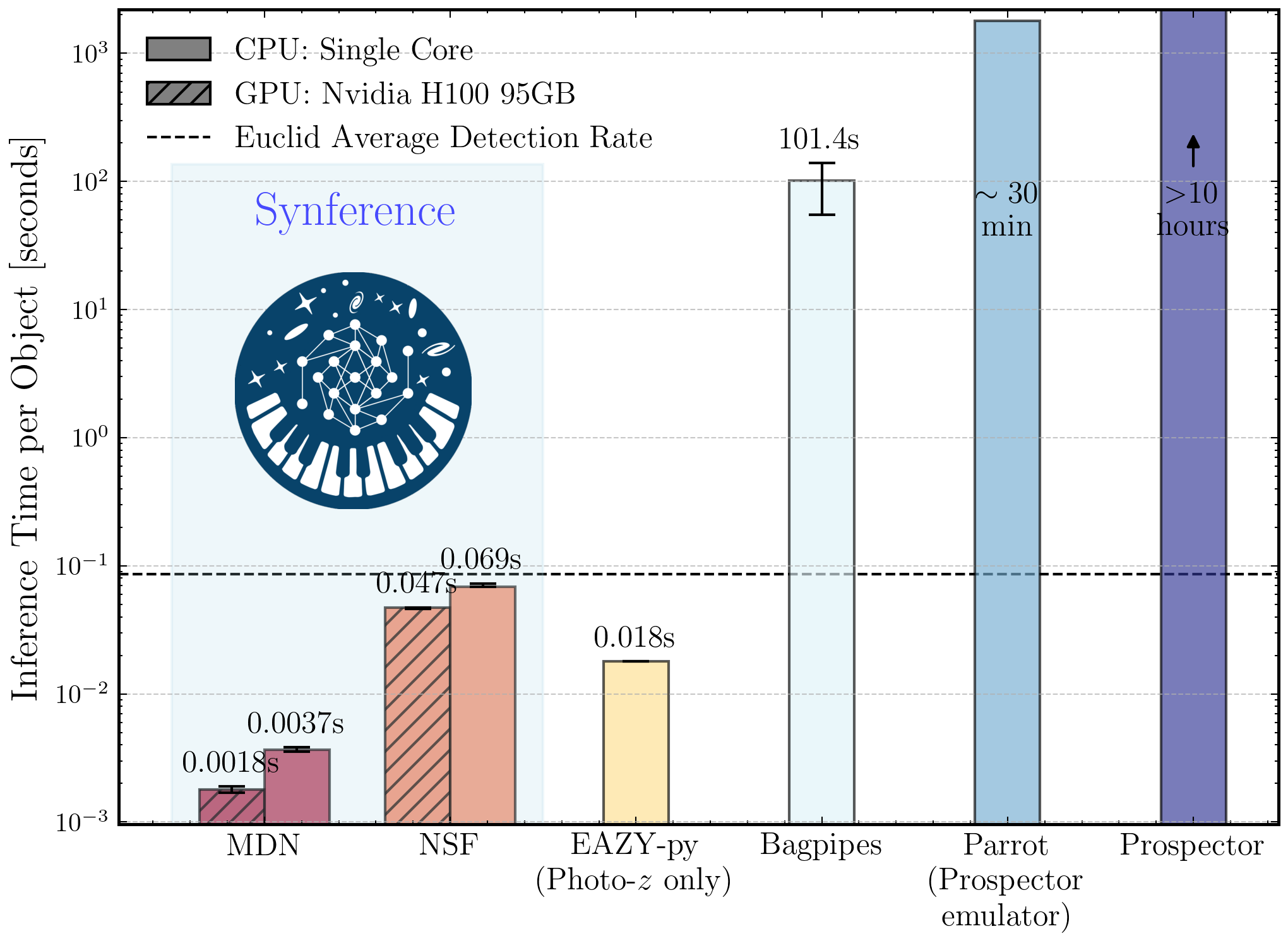}
    \caption{Comparison of inference speed with \synference{} (using \modelone{}, and a MDN variant of \modelone{}) for different hardware and density estimator choices to other SED fitting tools such as \eazy{}, \bagpipes{} and \prospector{}, run on the same galaxy sample. For CPU based inference we show single core performance to ensure a fair comparison. We also show the approximate speed of the Prospector emulator Parrot, from \protect\citep{mathews2023simple, wangUNCOVERSurveyFirstLook2023}. The horizontal dashed line shows the inference speed required to match the average speed at which galaxies are being observed across Euclid's 6 year mission, if processed in serial concurrently with detection \protect\citep{Euclid_overview}. Whilst this is purely illustrative it provides a visual example of the computational challenge of processing such large surveys. We do not include the training data generation or model training time here, since this only has to be performed once for a given configuration, and is not necessary during inference.}
    \label{fig:speed}
\end{figure}

\autoref{fig:speed} shows a comparison of the inference speed of our example models with \synference{} compared to other commonly used SED fitting tools. For \synference{}, which can utilize both the CPU and GPU for inference, we plot them separately. We note that GPUs do not typically offer a large speedup compared to CPUs for SBI models as the underlying neural networks within the NDE are relatively shallow. GPU performance will provide the largest speed-up if an embedding network with convolutional components is utilized alongside the SBI model for e.g. dimensionality reduction. 

This figure demonstrates one of the key strengths of the SBI approach, which is the speed of amortized inference. Depending on the neural network architecture we can perform inference at $20-100$ galaxies/second per galaxy using \synference{}, which is $>3$ orders of magnitude faster than with \bagpipes{} or \prospector{}. Whilst neural network emulators like Parrot are also very powerful when used to speed up likelihood evaluations in SED fitting, utilizing them in this fashion is still significantly slower than direct NPE, which altogether bypasses likelihood evaluations entirely. 

Purely as a demonstrative example \autoref{fig:speed} shows the inference speed required to match the average speed at which galaxies are being observed across Euclid’s 6 year mission, if processed in serial concurrently with detection \citep{Euclid_overview}. Whilst unrealistic, this is purely to demonstrate the volume of new galaxy observations which are currently being observed, and the scale of the data which will soon be available, if are able to make use of it. 

The speed of inference of the SBI model can depend on the observational data used for inference, and can be reduced significantly if the model is misspecified relative to the observations, or is simply not well optimized in general. This is because in the SBI framework whilst we employ prior constraints on our parameter sampling space, these constraints are not explicitly known by the neural network density estimator, which can produce samples outside of the parameter bounds, which is known as `leakage'. The prior constraints are instead utilized for accept-reject sampling when determining whether all parameter draws in a given posterior sample are within the prior bounds. How often a model produces samples outside the prior volume can significantly impact the sampling speed and cause low parameter acceptance. This problem is multiplied when sampling models with high dimensionality, as a rejected value for any dimension will reject the entire posterior sample. This can reduce the sampling speed for individual objects by several orders of magnitude relative to normal inference speed, and sometimes lead to the inference getting `stuck', with low or essentially zero accepted samples. In \synference{} this is handled by setting a maximum sampling time ($10-30$s) per galaxy, beyond which samples are instead sampled directly from the prior distribution. 

\subsection{Redshift Inference}

Spectroscopic redshifts are not typically available for the majority of galaxies in a photometric survey, especially in wide area photometric surveys (e.g. Euclid, LSST, Roman) where SBI techniques are particularly useful due to the speed of inference. In this case we can either rely on external photometric redshift estimates, e.g. template fitting with \eazy{} \citep{brammer2008eazy} or ML-based photometric redshifts \citep{2003MNRAS.339.1195F,2004PASP..116..345C,2019NatAs...3..212S,newman2022photometric}, or we can treat the redshift as unknown during inference and directly infer the redshift posterior from the observations. We train a version of our model with the same training data and hyperparameters, where we do not provide the redshift as a feature but instead train the network to learn the redshift. We refer to this as \modeltwo.

\begin{figure}
    \centering
    \includegraphics[width=\linewidth]{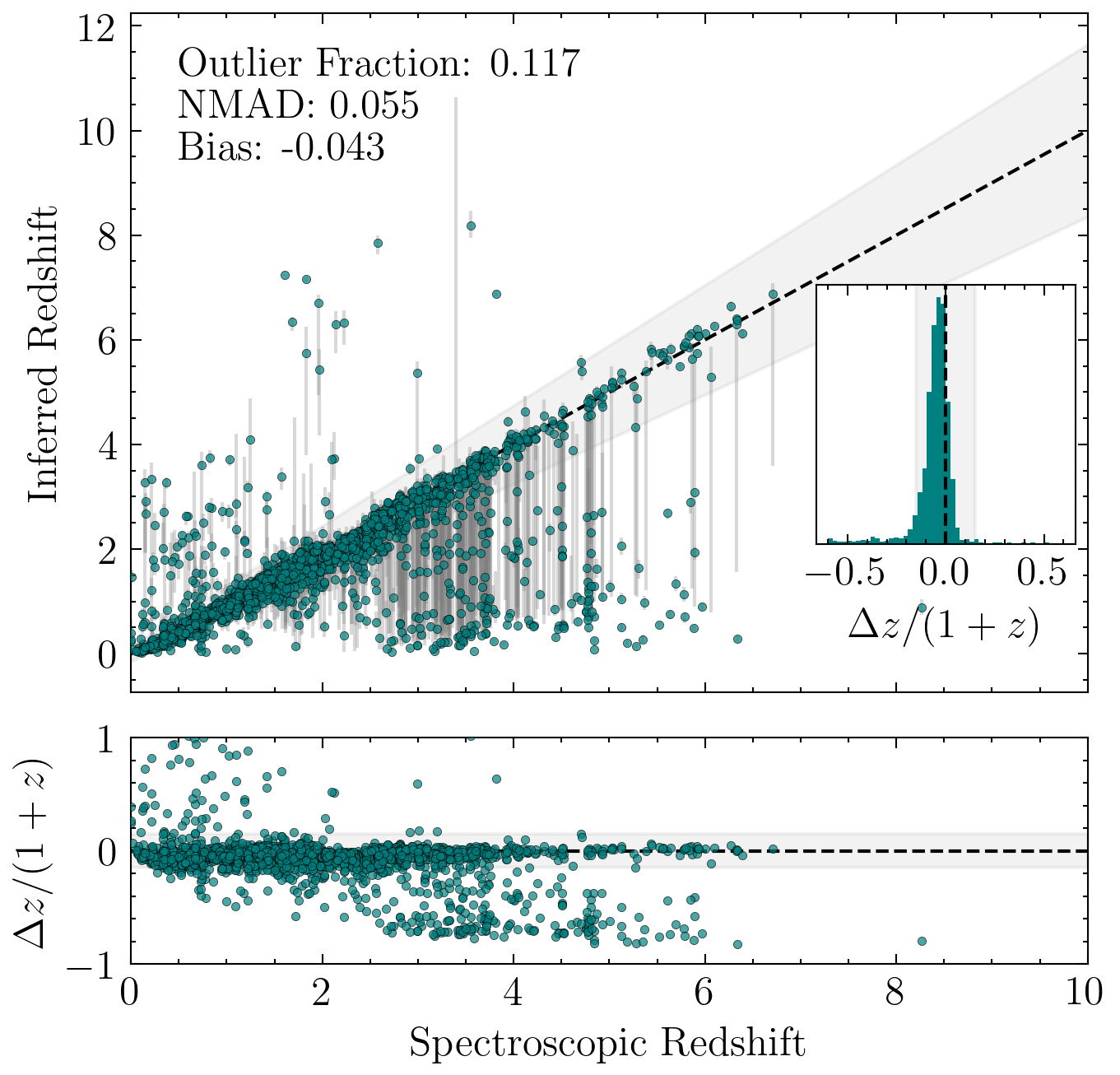}
    \caption{Photometric vs spectroscopic redshift for \modeltwo{} for the JADES spectroscopic catalogue, with the lower panel showing the normalized $\Delta z/(1+z_s)$ residuals, and an inset panel showing a histogram of $\Delta z/(1+z_s)$. \modeltwo{} is slightly biased towards lower redshifts (Bias of -0.043), which may be due to inflexibility in the training dataset (Ly$\alpha$ emission, DLAs, emission line ratios). The outlier fraction for this sample is 11.7\%, many of which are due to Lyman-Balmer confusion.}
    \label{fig:redshift}
\end{figure}

In \autoref{fig:redshift} we show the spec-$z$ vs. photo-$z$ comparison for the spectroscopic galaxy sample. We find an outlier fraction ($|\Delta z|/(1+z) > 0.15$) of $\eta=0.117$, a normalized median absolute deviation (NMAD) of 0.055 and a bias of -0.043. The lower panel shows $\Delta z/(1+z)$ as a function of the spectroscopic redshift. We observe that the primary outlier source is Lyman-Balmer break confusion, and that overall we see systematically underestimated redshifts, particularly in the $0 \leq z \leq 3$ redshift range, which can be seen in the inset histogram. This may be due to a lack of flexibility in our IGM prescription, not accounting for Ly-$\alpha$ emission correctly, or a lack of flexibility in the range of emission line ratios due to our fixed ionisation parameter and gas-phase metallicity. 

For comparison the \eazy{} photometric redshifts for the same sample had a outlier fraction $\eta=0.160$, an NMAD of 0.04, and a bias of -0.005. The \eazy{} model produces more outliers than \synference{}, but the redshift estimates are less biased with \eazy{}. It is encouraging to see that \modeltwo{} can recover accurate redshifts in the vast majority of cases. Often in the cases where the incorrect photo-$z$ is found, the redshift PDF is bimodal. A more complex model with a variable dust attenuation law slope, 2175\AA\ bump, or variable IGM prescription may shift more of the posterior weight towards the correct solution. 

In terms of computation speed, utilizing SBI techniques for only redshift inference is still slower than using \eazy{} in most scenarios. It is somewhat difficult to do a direct comparison between \eazy{} and our SBI model, given the difference in hardware requirements. For fairness our \eazy{} metrics are computed on a single CPU core, but we caution that in reality \eazy{} is likely to be parallelized over multiple cores which will further increase its performance, so this should be taken as a lower limit. However, we can similarly parallelise our inference over multiple cores by copying the trained \synference{} model, and many instances of the trained model could be used in parallel for inference across a large dataset. As shown in \autoref{fig:speed}, our performance metrics show that \eazy{} falls between the speed of the NSF and MDN models, at $\geq 50$~galaxies/second/core. The strength of the SBI approach to SED fitting is not the direct comparison to \eazy{} for just photo-$z$ estimation, but the ability to infer accurate posteriors for the physical parameters of the galaxies. Whilst \eazy{} does have some ability to compute estimates of parameters such as stellar mass, SFR, and stellar age based on best-fitting template weights and normalizations, this is not done in a Bayesian framework and does not produce Bayesian posteriors and covariances. A more reasonable speed comparison for \synference{} is Bayesian photometric redshift estimation alongside parameter estimation, e.g. Prospector-$\beta$ \citep{wang2023inferring} or \bagpipes{} with a uniform redshift prior. Prospector-$\beta$ takes >10 hours per fit with nested sampling \citep{wang2023inferring}, and when emulated with \textsc{Parrot} \citep{mathews2023simple} this can be reduced to $\sim$30 minutes/galaxy. \bagpipes{} can take $\sim1-5$ minutes per galaxy to fit using nested sampling, but is somewhat less flexible than \prospector{}. \synference{} shows a $10^{3}-10^{5}$ speed up over the traditional Bayesian approaches to redshift + galaxy parameter inference.

\subsection{Model Comparison}

SED fitting codes are typically built with some limited flexibility around a fundamental set of fixed assumptions. For example, SFH parametrizations and dust attenuation models can often be varied, whilst more fundamental assumptions such as the underlying stellar population model or initial mass function are more difficult to change. With \synference{} the user can be more flexible, as most of the major stellar population synthesis models are supported through \synthesizer{}, along with a variety of IMF parametrizations. Grids with higher dimensionality than just stellar age and metallicity, allowing for direct interpolation across e.g. the IMF high-mass slope, are also available. Users also have the ability to easily produce and post-process their own custom SPS grids if needed\footnote{https://github.com/synthesizer-project/grid-generation}. 

\begin{figure}
    \centering
    \includegraphics[width=\columnwidth]{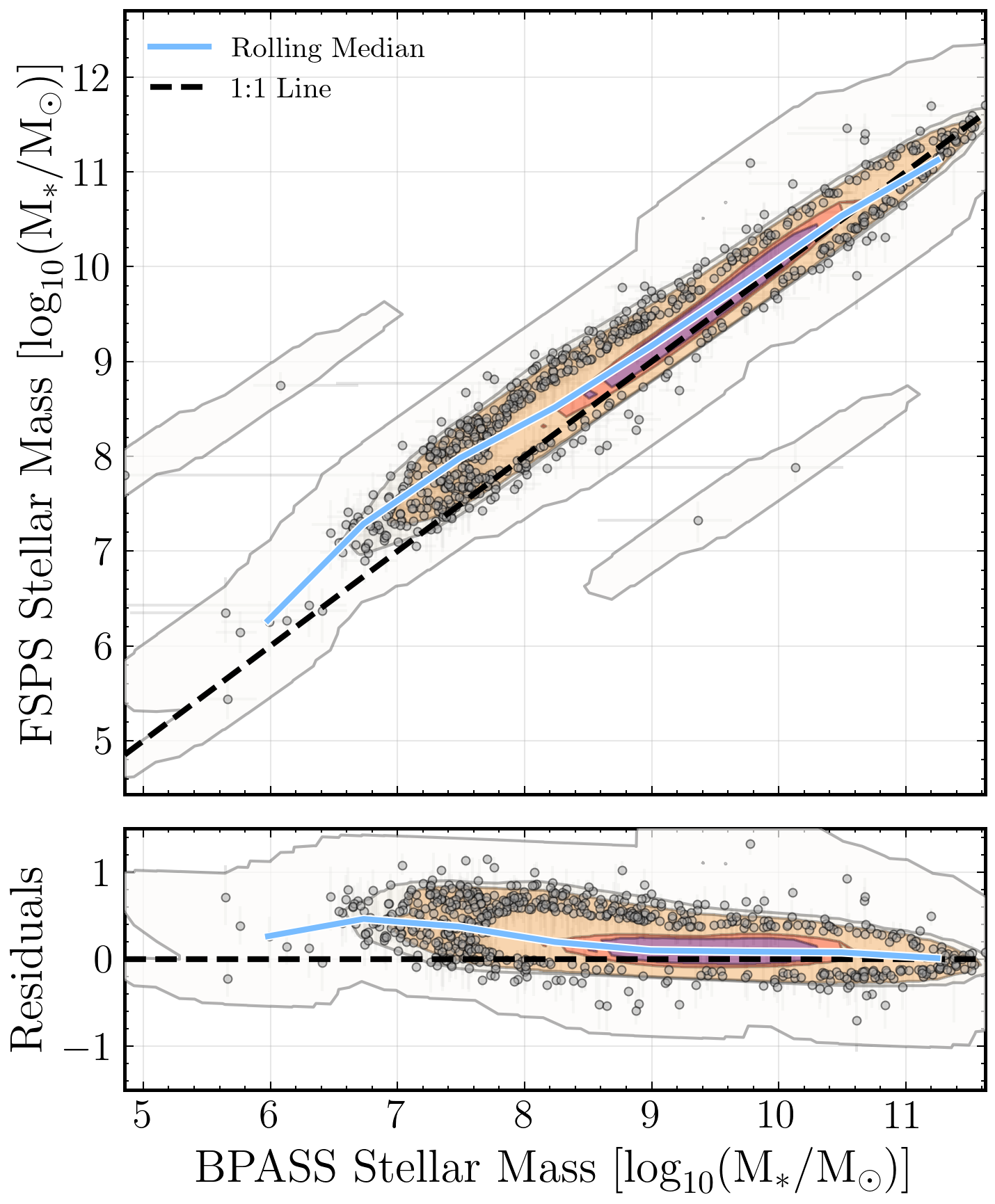}
    \caption{Stellar mass comparison between \synference{} models trained on photometry generated from BPASS v2.2.1 SPS grid (\modelone{}) and a new model trained on photometry from a SPS grid FSPS grids \protect\cite{conroy2010fsps}. Both grids use the same \protect\cite{Chabrier2000} IMF. The FSPS grid produces systematically higher stellar mass estimates, especially at $\log_{10}\rm M_\star^{BPASS} < 8$, with a median offset of 0.3~dex. Most points are contained within the density contours, and only outliers below the 20th percentile of KDE density are plotted as individual points with errorbars.}
    \label{fig:comparison}
\end{figure}

This flexibility, along with the speed at which a new model can be trained and used for inference, allows for model selection and comparison, which is difficult to accomplish with a traditional SED fitting approach. As a demonstration of this we have trained a new version of our final model with the same hyperparameters and network architecture, but where we have replaced our BPASS \citep{stanway2018re} SPS grid with an FSPS \citep{conroy2010fsps} grid, both with the same \cite{Chabrier2000} IMF and post-processed in the same way. In \autoref{fig:comparison} we show a comparison of the inferred stellar mass between the model using the BPASS SPS grid and the new model constructed from the FSPS grid. 

Interestingly, we see that with the same IMF, the FSPS grid produces systematically larger stellar mass estimates. At $\log_{10}\rm M_\star^{BPASS} < 8$ this results in a median offset of 0.3~dex (2$\times$). As both models use the same underlying \cite{Chabrier2000} IMF, this observed mass offset is likely due to differences in the M/L ratios between the two SPS models. 

SBI can also be used for explicit Bayesian model comparison and selection, e.g. using evidence ratios (the Bayes factor). There have been a number of studies implementing model comparison within the SBI framework, including Evidence Networks \citep{jeffreyEvidenceNetworksSimple2024} and Harmonic Evidence \citep{harmonic, spuriomanciniBayesianModelComparison2023}. We plan to explore these approaches within \synference{} in future works.

\subsection{Comparison to Existing Tools}

Whilst the application of SBI to SED fitting is relatively new, there have been a few other implementations which we can compare with \synference{}. This includes \textsc{SBI++} \citep{wangSBIFlexibleUltrafast2023} and \textsc{sbipix} \citep{iglesias-navarroSimulationbasedInferenceGalaxy2025}. 
Both of these approaches have been shown to perform well, with \citep{wangSBIFlexibleUltrafast2023} robustly handling missing data and outliers, and \citep{iglesias-navarroSimulationbasedInferenceGalaxy2025} optimized for rapid pixel level inference based on a pre-determined model. \synference{} is designed to act as more of a framework, with greater flexibility in the freedom the user has to build and train their model, whereas the existing tools implement a specific SED model and training architecture. \synference{} also enables additional functionality such as hyperparameter optimization, built-in model validation and calibration tests, and SED recovery. 

\subsection{Limitations}

Here we address a few of the limitations of the current \synference{} approach. Firstly, there are still some minor inflexibilities in the \synthesizer{} forward model which will be improved in the future. These include e.g. full support for high-dimensional SPS grids, which will allow e.g. variation in ionisation parameter. This would allow greater flexibility in emission line ratios and strengths for star-forming galaxies. 

Overall model performance is another area where improvements may be possible. As discussed in \autoref{sec:performance}, some parameters, particularly those which are covariant, such as the overall SFH, are not recovered as precisely as traditional SED fitting approaches. Whilst our large hyperparameter optimization has located the optimal configuration for our model, there may be other neural density estimators, approaches such as Neural Score Posterior Estimation \citep[][]{sharrock2022sequential} or transformer approaches such as the \textsc{simformer} \citep{gloecklerAllinoneSimulationbasedInference2024}, which may allow us to further improve model performance. 

As discussed in \autoref{sec:sbi_bckg}, missing data is an ongoing challenge in applying SBI techniques to real data. All results presented here have utilized only observations where measurements in all filters are available. We have incorporated multiple approaches to dealing with missing data into \synference{}, but we leave a detailed discussion and exploration of these approaches to a future work.


\section{Summary and Future Work}
\label{sec:conclusions}

In this paper we have described \synference{}, a new tool for SED fitting using Simulation-Based Inference. We have demonstrated the performance of \synference{} on a example model, and validated the model on simulated and observed photometry from the JADES GOODS-South field. This tool is built to address the significant computational challenge posed by the immense datasets from current and future surveys like JWST, Roman, and Euclid, for which traditional inference methods are prohibitively slow.\par\bigskip\noindent
Our main results are summarized as follows:
\begin{enumerate}
    \item \noindent \textbf{A New Flexible Framework}: \synference{} provides a modular and extensible SBI tool for SED fitting. It decouples data simulation (using the flexible \synthesizer{} package) from neural network training (using the \ltuili{} backend), allowing a single library of simulations to be reused for multiple science applications. Training data can also be generated externally e.g. from forward modelled outputs from hydrodynamical simulations. \par\bigskip

   \item \noindent \textbf{Robust Model Training and Validation}: We detailed our process for feature engineering, empirical noise modelling, and hyperparameter optimisation using \textsc{Optuna}, which identified a Neural Spline Flow (NSF) as the optimal architecture for our problem. We rigorously validated our fiducial 8-parameter model, \modelone{}, against a held-out test set and reference posteriors from \textsc{dynesty}. The model demonstrates excellent parameter recovery ($R^2=0.99$ for $\rm M_\star$, $R^2=0.86$ for $\rm A_V$) and reliable posterior calibration, passing all TARP and coverage tests.\par\bigskip

    \item \noindent \textbf{Exceptional Inference Speed}: We applied \modelone{} to a sample of 3,088 galaxies with spectroscopic redshifts in the JADES GOODS-South field. The amortized inference for the entire sample took $\sim$ minutes on a desktop CPU, a speedup of $\sim$1700$\times$ compared to the $\sim$80 CPU-hours required for an equivalent analysis with the SED fitting code \bagpipes{}.\par\bigskip

    \item \noindent \textbf{Accurate Application to Real Data}: The physical properties inferred by \synference{} show good agreement with those from \bagpipes{}. We found that \synference{} was more robust in fitting quiescent galaxies where \bagpipes{} struggled to find the correct solution, highlighting the ability of NPE to handle complex posterior landscapes.\par\bigskip

    \item \noindent \textbf{Expanded Capabilities}: We demonstrated the power of \synference{} for applications beyond standard SED fitting. By training a model to infer redshift (\modeltwo{}), we showed it can produce full Bayesian posteriors for both redshift and physical parameters at a speed $10^{3}-10^{5}\times$ faster than traditional Bayesian methods. We also highlighted its use for rapid model comparison by training a new model with a the FSPS SPS model (instead of BPASS) and quantifying the systematic 0.3 dex offset in stellar mass.\par\bigskip
\end{enumerate}

\noindent \synference{} provides a scalable and powerful solution for exploiting the next generation of astronomical surveys, enabling fast, full-posterior inference for billions of galaxies. We plan to apply \synference{} in numerous contexts in future work, including in spatially resolved galaxy studies, inference from hydrodynamical simulations, Bayesian model comparison, and high-dimensional inference from spectroscopy.

\section*{Acknowledgements}

TH, CC, NA, DA, QL, VR, KM, acknowledge support from the ERC Advanced Investigator Grant EPOCHS (788113), as well as two studentships from the STFC.
CCL was supported by the research environment and infrastructure of the Handley Lab at the University of Cambridge.
WJR, APV and SMW acknowledge support from the Sussex Astronomy Centre STFC Consolidated Grant (ST/X001040/1). PIN thanks the LSST-DA Data Science Fellowship Program, which is funded by LSST-DA, the Brinson Foundation, the WoodNext Foundation, and the Research Corporation for Science Advancement Foundation; her participation in the program has benefited this work. PIN and MHC acknowledge financial support from the State Research Agency of the Spanish Ministry of Science and Innovation (AEI-MCINN) under the grants ``Galaxy Evolution with Artificial Intelligence'' with reference PGC2018-100852-A-I00 and ``BASALT'' with reference PID2021-126838NB-I00. All contributions from NA were made before he left academia. 

This work is based on observations made with the NASA/ESA \textit{Hubble Space Telescope} (HST) and NASA/ESA/CSA \textit{James Webb Space Telescope} (JWST) obtained from the \textsc{Mikulski Archive for Space Telescopes} (\textsc{MAST}) at the \textit{Space Telescope Science Institute} (STScI), which is operated by the Association of Universities for Research in Astronomy, Inc., under NASA contract NAS 5-03127 for JWST, and NAS 5–26555 for HST.

This work used the DiRAC@Durham facility managed by the Institute for Computational Cosmology on behalf of the STFC DiRAC HPC Facility (www.dirac.ac.uk). The equipment was funded by BEIS capital funding via STFC capital grants ST/P002293/1, ST/R002371/1 and ST/S002502/1, Durham University and STFC operations grant ST/R000832/1. DiRAC is part of the National e-Infrastructure.

\section*{Data Availability}

The \textsc{synference} package used to generate the training data and train the SBI models is available on  GitHub at \url{https://github.com/synthesizer-project/synference}. Documentation for using \synference{} is available at \url{https://synthesizer-project.github.io/synference}. \synthesizer{} is also available on GitHub at \url{https://github.com/synthesizer-project/synthesizer}. \modelone{} is available to download through the \texttt{synference-download} tool built into the package, and we plan to distribute more pre-trained models in future works.

The EPOCHS JADES catalogue and imaging are available upon reasonable request to the corresponding author, and will be made public as part of the EPOCHS-v2 project (Austin et al. in prep.).



\bibliographystyle{mnras}
\bibliography{main} 




\appendix

\section{Accuracy of Bagpipes Inference}

In \autoref{sec:gal_inference} we compare galaxy properties inferred with \synference{} to the results of the standard \bagpipes{} Bayesian SED fitting tool. However given that this comparison is done on a catalogue of real galaxies, rather than mock data, we do not have access to the ground truth to assess the absolute performance of either tool. Whilst we assess the accuracy and reliability of \synference{} in \autoref{sec:validation}, we do not validate whether \bagpipes{} itself can accurately recover model parameters. In this appendix we test \bagpipes{} using the same validation dataset we use for \synference{}. 

\begin{figure*}
    \centering
    \includegraphics[width=0.8\textwidth]{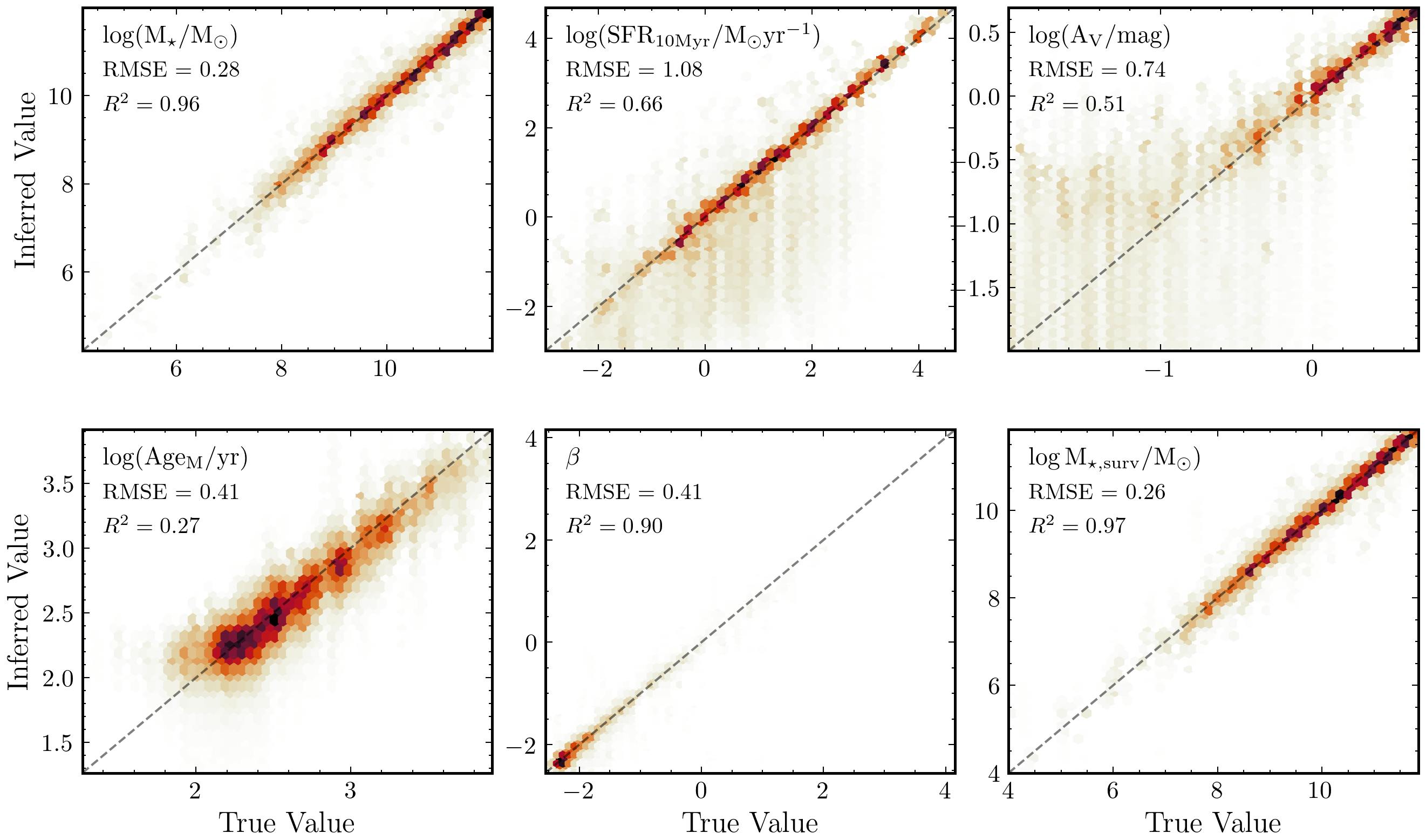}
    \caption{Equivalent of \autoref{fig:model_validation} for \bagpipes{}, showing model performance on $\geq 5\sigma$ validation set. }
    \label{fig:bagpipes_validation}
\end{figure*}

We randomly draw 1000 galaxies from the subset of the validation dataset for \modelone{} which has average SNR $>5$ in F277W, F356W and F444W. We fit these galaxies using \bagpipes{} in the same way as described in \autoref{sec:gal_inference}. We compute $R^2$ and RMSE in the same method as for \synference{}.

\autoref{fig:bagpipes_validation} shows the equivalent of \autoref{fig:model_validation} for the equivalent \bagpipes{} results as determined by comparing the ground-truth of the validation dataset to the results of \bagpipes{} fitting. The \bagpipes{} model is less accurate and more biased than the \synference{} model. The figure shows the $R^2$ and RMSE values for each parameter. The RMSE errors for these parameters using \bagpipes{} are lower than or comparable to  \modelone{} for the same subset of the validation dataset, which can be compared with \autoref{tab:model_performance}. This suggests that \synference{} can produce results which are more accurate than \bagpipes{}.

\bsp	
\label{lastpage}
\end{document}